\documentstyle[12pt,aasms4,psfig]{article}

\newcommand{\solm}{M$_{\odot}$}

\newcommand{\COz}{$^{12}$CO (2-1) }
\newcommand{\COe}{$^{12}$CO (1-0) }
\newcommand{\CO}{$^{12}$CO }

\newcommand{\rf}{\par\noindent\hangindent 15pt {}}
\newcommand{\apjj}[2]{Ap. J., #1, #2.}
\newcommand{\apjjs}[2]{Ap. J. Supp., #1, #2.}

\newcommand{\asa}[2]{Astron. Astrophys., #1, #2.}
\newcommand{\arasa}[2]{Ann.Rev.Astr.Ap., #1, #2.}
\newcommand{\asas}[2]{Astron. Astrophys. Suppl., #1, #2.}
\newcommand{\aapss}[2]{Astrophys. and Space Science #1, #2.}

\newcommand{\mn}[2]{M.N.R.A.S., #1, #2.}
\newcommand{\ajj}[2]{A. J., #1, #2.}

\newcommand{\vol}[1]{1}
\newcommand{\nhico}{$\frac{N_{H_2}}{I_{CO}}$}
\newcommand{\htwo}{$H_2$ }

\lefthead{Schinnerer et al.}
\righthead{Molecular Gas in NGC~3227}

\begin{document}

\title{Distribution and Kinematics of the Circum-nuclear 
       Molecular Gas in the Seyfert~1 Galaxy NGC~3227}

\author{
E. Schinnerer\altaffilmark{1}, A. Eckart, L.J. Tacconi }

\affil{Max-Planck-Institut f\"ur extraterrestrische Physik, 85740 Garching, Germany}

\altaffiltext{1}{Present address: California Institute of Technology, 
Pasadena, CA 91125}

\altaffiltext{2}{Ap. J. accepted}

\begin{abstract}
We present new interferometric observations of the 
\COe, \COz, and HCN (1-0) molecular line emission in NGC~3227 
obtained with the IRAM Plateau de Bure 
interferometer (PdBI). We achieved an 
unprecedented angular resolution in the \COz line of
about 0.6'' corresponding to only about 80~pc at a distance of 17.3~Mpc. 
The mapped \CO ~emission is concentrated
in the inner 8'' and accounts for 20\% of the total 30~m  \CO line flux. 
The \CO emission
is resolved into an asymmetric nuclear ring with a diameter of about 3''.
The HCN line emission is mostly unresolved at our resolution of
$\sim$ 2.4'' and contains all of the single dish flux. 
We have decomposed the observed molecular gas motions into a circular
and non-circular component revealing that about 80~\% of the gas in the 
circum-nuclear region exhibits pure circular rotation. We find
evidence for bar streaming onto the nuclear ring and a
redshifted emission knot on the ring perimeter. 
\\
In the central arcsecond the gas shows
apparent counter rotation. This behavior
can be best explained by a warping of the inner molecular gas disk
rather than gas motion in a nuclear bar potential.
We detected molecular gas at a distance
from  the nucleus
of only $\sim$13~pc
with a velocity of about 75 km/s with respect to the systemic velocity
and find that within the central arcsecond  the rotation curve is
rising again.
This is the first time that millimetric CO-line emission has been detected
interferometrically at such small distances to the nucleus of a Seyfert
galaxy. 
These measurements indicate  a lower limit
on the enclosed mass of about 2$\times$10$^7$ \solm ~in the inner 25~pc.
\end{abstract}


\section{INTRODUCTION}

The distribution of circum-nuclear molecular gas plays an important
role in the proposed unified scheme for Seyfert galaxies. In the
standard picture a torus of dense molecular gas and dust is
surrounding the black hole and its accretion disk 
(e.g. overviews by Antonucci 1993 and Peterson 1997, see also
Pier \& Krolik 1992, 1993, Edelson \& Malkan 1987).
 As the relative orientation of the AGN to
the plane of its host galaxy is probably random the appearance of the AGN as
Seyfert~1 or Seyfert~2 nucleus depends only on whether the viewing angle
onto the central engine is blocked by the torus or not.
There are two important questions arising for the distribution and
kinematics of the circum-nuclear molecular gas: 
\\
(1) Is the circum-nuclear molecular gas already participating in the
obscuration of the AGN? 
Recent observations of about 250 nearby active
galaxies with the HST from Malkan, Gorjian \& Tam (1998) suggest
that molecular material at distances of about 100~pc is responsible
for the obscuration of the nucleus rather than a nuclear torus.
Similarly Cameron et al. (1993) and Schinnerer, Eckart \& Tacconi (1999)
find indications for a more complex
picture in NGC~1068.
\\
(2) What is the fueling mechanism for AGNs? As a very effective
mechanism nuclear bars are considered to bring the molecular gas to small
radii and fuel the central engine
(e.g. Shlosman, Frank \& Begelman 1989). However, Regan \& Mulchaey
(1999) searched in 12 nearby Seyfert galaxies with little success 
for signatures of strong
nuclear bars by combining HST NICMOS 1.6 $\mu$m images with HST
optical images to study the dust morphology (see also Martini \& Pogge 1999). 
\\
\\
The recent improvements in mm-interferometry now allow one to obtain
sub-arcsecond resolution observations of the molecular line emission
in combination with high spectral resolution and high sensitivity. This 
combination is ideal to study the
kinematics and distribution of molecular gas in the circum-nuclear
region of nearby Seyfert galaxies.
\\
\\
NGC3227 (Arp~94b) is a Seyfert 1 galaxy located at a distance of 17.3~Mpc
(group distance; Garcia 1993; see also Tab. \ref{ww1}).
Rubin \& Ford (1968) studied the system NGC~3226/7 for the first time in the 
optical. In NGC~3227 they found indications for a nuclear outflow 
as well as a spiral arm that stretches toward the close elliptical companion.
The NLR is extended towards the northeast (Mundell et al. 1992a,
Schmitt \& Kinney 1996) and the BLR clearly shows variations in the 
optical continuum and line emission (Salamanca et al. 1994, Winge et al. 1995).
\\
NGC~3227 is classified as a SAB(s) pec in the RC3 catalog (de Vaucouleurs et 
al. 1991). Due to its
inclination the end of the bar and the starting points of the spiral
arms are difficult to identify. 
As indicated by optical and NIR observations (Mulchaey, Regan, Kundu 1997, De
Robertis et al. 1998) the galaxy has probably a bar with 6.7~kpc -
8.4~kpc radius and a position angle of about -20$^o$ relative to the major
kinematic axis. 
Mundell et al. (1995b) studied the system NGC~3226/7 in the
HI line emission. Two gas plumes north and south of the system
are seen with velocities between about 1200 km/s and 1300 km/s. The HI
gas of the plumes with the largest spatial separation from NGC~3227
also shows the largest velocity difference to NGC~3227
($v_{sys}$ = 1110$\pm$10~km/s; derived in section \ref{ww12}). These clumps
can be identified with tidal tails which are often observed in 
interacting galaxies. 
\\
\\
Here we mainly study the distribution and kinematics of the
molecular gas in the central few arcseconds of NGC~3227. The
observations are summarized in section \ref{ww01} followed by the description
of the interferometric data (section \ref{ww04}). 
In section \ref{ww12} we outline molecular gas kinematics including the 
derivation of the rotation curve and
a discussion of the general structural and dynamical properties.
A decomposition of the
molecular gas in NGC~3227 into components of circular and non-circular
motion is given in section \ref{ww24} (and Appendix A). 
The mass and thickness of the molecular gas disk are derived in section
\ref{ww28}. In section \ref{ww30} (and Appendix B and C) 
a 3-dimensional analysis of the kinematics in the inner 1'' (85~pc)
in NGC~3227 in the framework of a bar and warp model for the gas
motion is presented. The final summary and implications
are given in section \ref{ww34}.

\section{\label{ww01}OBSERVATIONS AND DATA REDUCTION}

The \COe and \COz lines were observed in January/February 1997 using the
IRAM PdB interferometer with 5 antennas in its A, B1 and B2 configuration 
providing 30 baselines from 40~m to 408~m. The resulting resolution is
1.5'' $\times$ 0.9'' (PA 34$^o$) at 2.6~mm and 0.7'' $\times$ 0.5''
(PA 31$^o$) at 1.3~mm using uniform weighting. 
We applied standard data reduction and calibration
(15\% accuracy at 2.6 and 1.3~mm)  procedures.
After cleaning we reconvolved data with a
CLEAN beam of FWHM 1.2'' for \COe and FWHM 0.6'' for the \COz.
To increase the S/N we smoothed the data
spectrally to a resolution of 20 km/s in both lines. In addition a \COz data
set with a spectral resolution of 7 km/s was made ranging
from -182 km/s to 182 km/s since only this interval was covered by
all three configurations.
\\
The HCN (1-0) line was observed between March 1995 and March 1996
using the IRAM PdB interferometer with 4 antennas in its B1, B2, C1
and C2 configuration providing baselines from 24~m to 288~m. The spatial
resolution is 3.3'' $\times$ 1.7'' (PA 42$^o$) using natural weighting. 
 The uncertainties of
the flux calibration are about 15 \%.
For the interferometric observations we used a velocity of
v$_{obs}$=1154~km/s which is 44 km/s larger than the systemic velocity
of v$_{sys}$=1110~km/s derived from the data (section \ref{ww12}).

\section{\label{ww04}RESULTS}

The IRAM 30~m data (Schinnerer et al. in prep.)
show an elongation of the nuclear
line emission at a PA of $\sim$130$^o$.
A comparison to published data reveals that even the 30~m telescope already
starts to spatially resolve parts of the central 2~kpc diameter emission.
At high angular resolution  most of the molecular gas is located
in a 3'' diameter circum-nuclear ring.
In the following subsections we describe the distribution and kinematics
of the molecular gas as observed with the PdBI in the 
$^{12}$CO(1-0), $^{12}$CO(2-1), and the HCN(1-0) line.
The detection of molecular line emission at radii as small as 13~pc is
used to estimate an upper limit to the enclosed nuclear mass. No millimeter
radio continuum emission was detected by our interferometric
observations. The corresponding upper limits are
8~mJy at 1.3~mm and 5~mJy at 2.6~mm.

\subsection{\label{ww07}The extended emission}

\begin{figure}
\begin{center}
\psfig{file=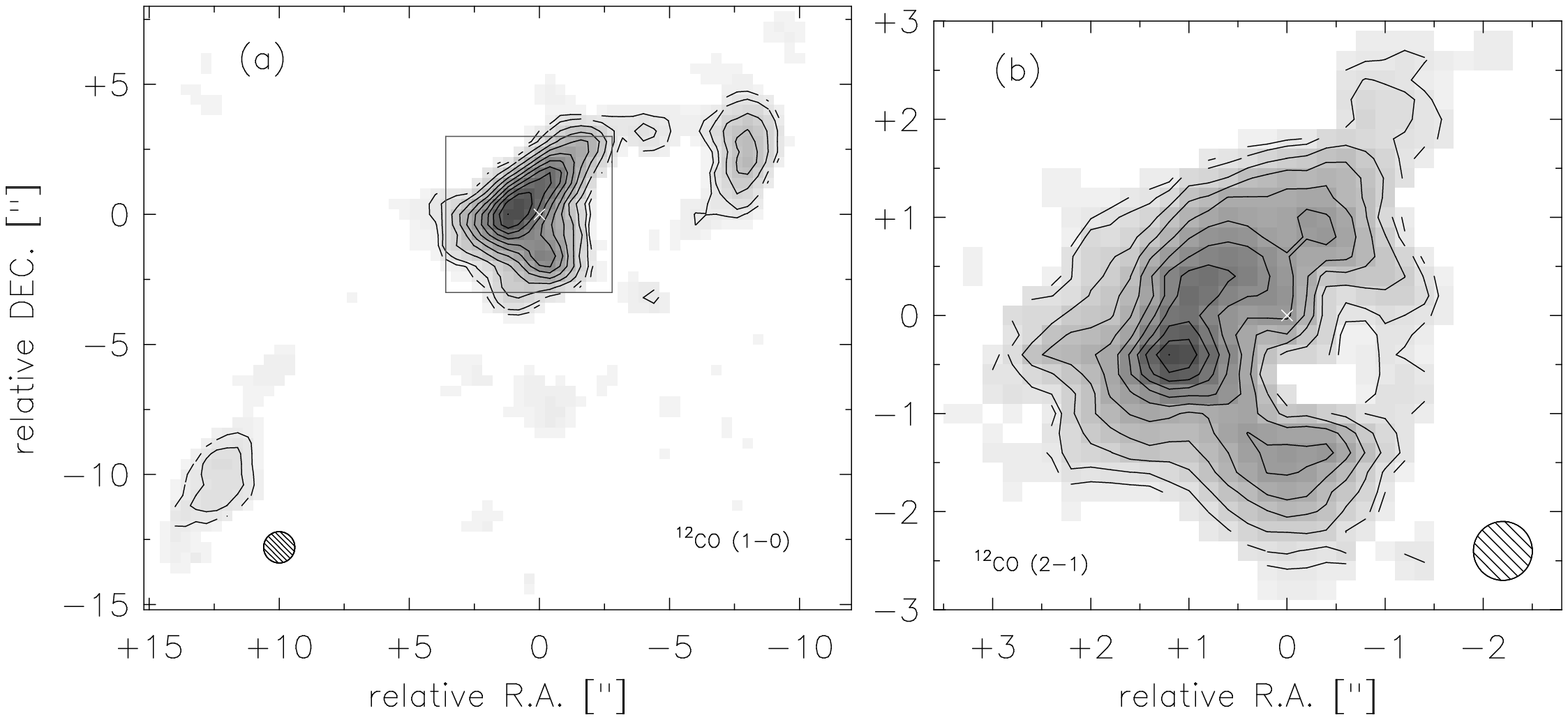,height=10.0cm,width=14.1cm,angle=-90.0}
\end{center}
\figcaption[]{
\label{abb02}
The IRAM PdBI maps of NGC~3227. The intensity maps (0$^{th}$-order moment)
 of the
\COe line emission (left) at a resolution of 1.2'' and in contours of
5, 10, 20, ... 100 \% of the maximum of 8.78 Jy/beam km/s. The
indicated area corresponds to the \COz map show in the right panel.
The contours of
the \COz map at a resolution of 0.6'' are in 5, 10, 20, ... 100\% of
the peak of 9.18 Jy/beam km/s. The \COe map contains only about 20\%
of the intensity of the IRAM 30~m map whereas the \COz map
contains about 10\% of the total line intensity of the 30~m map.}
\end{figure}

\begin{figure}
\begin{center}
\psfig{file=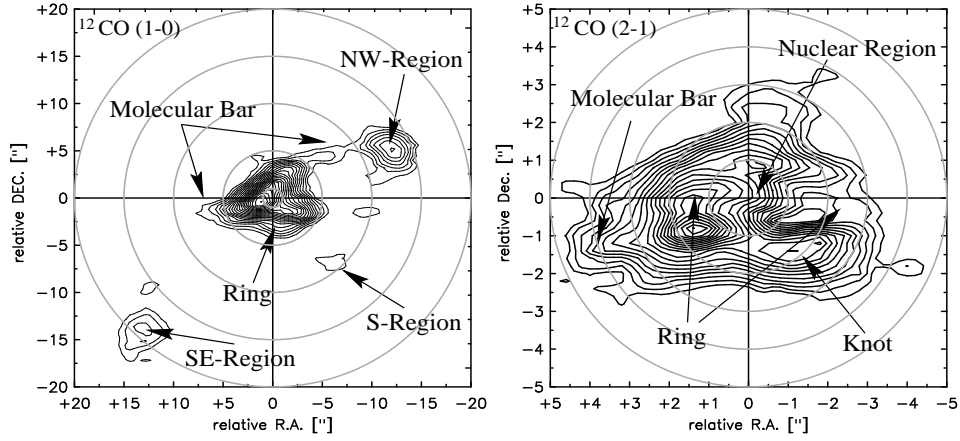,height=10.0cm,width=14.1cm,angle=-90.0}
\end{center}
\figcaption[]{
\label{abb03}
The de-projected maps of the PdBI \CO line emission. The maps are
de-projected by correcting for an inclination angle of $i$=56$^o$ and a
position angle of $PA$=158$^o$. To ease the comparison to the real
data the blue side of the major kinematic axis was aligned to the
north. The contours are 5, 10, 15, 20, ... 100\% of the peak
intensity. The regions discussed in the text are indicated.}
\end{figure}

The \CO emission mapped with the PdBI is concentrated in an uneven
ring-like structure in the inner 6'' (500 pc; Fig. \ref{abb02}). 
The eastern part is about six times brighter than the
western part (Fig. \ref{abb02}). 
In the following we refer to this structure as the {\it ring}.
In the channel maps \COe line emission is 
seen from -260 km/s to 220 km/s (relative to the central velocity
$v_{obs}$ = 1154~km/s we used for the observations) and 
from -240 km/s to 180 km/s for the \COz line at a spectral resolution of 20
km/s. The difference in velocity range is probably due to the fact that the 
smaller 0.6'' beam of the \COz line has already resolved some emission
which is still detected in the \COe beam of 1.2''. 
\\
\\
A comparison to single dish measurements 
shows that the PdBI \COe  and \COz maps contain about 20 \% and 10 \%
of the total line flux, respectively.
Combined with the results from the 30~m maps (Schinnerer et al. in prep.)
this means that most of the 
remaining \CO gas is distributed in a fairly smooth gas disk, and 
the structures seen in the PdBI maps are the main concentrations
of the molecular gas. 
However, the \CO line flux observed by the PdBI traces the compact 
circum-nuclear gas component that constitutes the reservoir out of which 
both the nucleus as well as circum-nuclear star formation can be fed.
\\
A much weaker additional component ({\it molecular bar}) is
detected east and west of the center. This component stretches out
to a radius of about 7'' (590 pc). The east and west parts of the bar have a 
NS-offset of about 3'' connecting the NW-region with the circum-nuclear ring
(Fig. \ref{abb02}). 
In the de-projected map this component resembles a bar
which encloses an angle of about 85$^o$ with the kinematic
major axis (Fig. \ref{abb03}).
For de-projection
the intensity maps were rotated by the position angle so that the
kinematic major axis is parallel to the $x$-axis. The $y$-axis was
then corrected via  $ y = y' / cos(i) $, where $i$ is the inclination.
\\
In addition emission regions to the north-west, south-east and south
({\it NW-region, SW-region, S-region}; Fig. \ref{abb03})
lie in a de-projected spatial distance of
10'' to 20'' from the center and have typical sizes of about
1.5'' to 2''. The NW- and SE-region are stronger in the \COe line
than in the \COz line whereas the S-region is more prominent in
the \COz line. This might indicate that the molecular line emission in
the S-region is partly due to optically thin gas.
The NW-region lies at the tip of the
molecular bar and is also twice as bright as the SE-region in the
\COe line emission.

\subsection{\label{ww10}The nuclear emission}

Over a velocity range from -140 km/s to 35 km/s
we detect emission with an extent $\sim$0.6'' at the dynamical 
center. 
In all pv-diagrams the structure of this component appears to be
{\sf S}-shaped and symmetric both with respect to the central position 
and with respect to a velocity about 
19 km/s below the systemic velocity 
$v_{sys; HI}$ = 1135~km/s
derived from HI observations by Mundell et al. 1995b (44 km/s 
below the central velocity of 
$v_{obs}$ = 1154~km/s we assumed for the observations). 
\\
\\
{\it The position of the dynamical center:}
To derive the exact position of the dynamical center we fitted a Gaussian to
the nuclear component in the channel map at -63 km/s.
We chose that map since its velocity is relatively close
to the true systemic velocity and the nuclear component is
clearly separated from the emission in the ring.
The nucleus is positioned (0.28$\pm$0.02)'' west and (0.84$\pm$0.03)''
north of the interferometer phase center of RA 10$^h$23$^m$30.590$^s$ 
and DEC 19$^o$51'54.00'' (J2000.0). 
We note that the position of the phase 
center may itself have an error of up to 0.2''.
With this uncertainty the positions of the northern and southern radio
component (Mundell et al. 1995b) are both included in our error budget.

\subsection{\label{ww11}The HCN (1-0) Data}

\begin{figure}
\begin{center}
\psfig{file=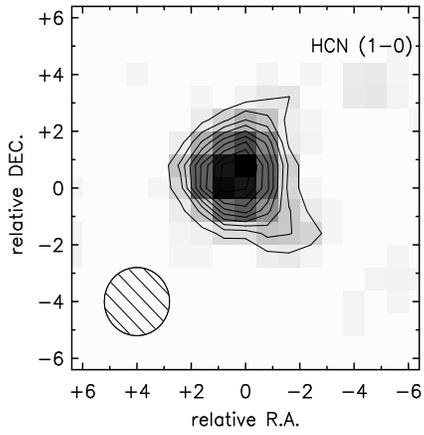,height=14.1cm,width=10.0cm,angle=0.0}
\end{center}
\figcaption[]{
\label{abb05}
IRAM PdBI map of the HCN(1-0) line emission in NGC~3227 at a resolution of
2.4''. The contours are 30, 40, ... 100\% of the peak of 1.93 Jy/beam
km/s.}
\end{figure}

The HCN (1-0) line emission in NGC~3227 was observed with the IRAM
30~m (Schinnerer et al. in prep.) as well as the PdBI (Fig. \ref{abb05}). The 
comparison of both measurements shows
that the PdBI map contains the total flux of the 30~m observation.
The HCN (1-0) line emission is concentrated in a barely
resolved nuclear source. A Gaussian fit provides
a source size of (2.67$\pm$0.24)'' very similar to the FWHM of the
beam of 2.4''. Therefore the HCN source structure is much more compact
than the \CO ring structure.
\\
The strongest HCN(1-0) emission (about twice as strong as in the
neighboring channel maps) is found at -40 km/s 
(close to the systemic velocity) indicating a strong nuclear component.
The position of the peak within its errors in this map is 
identical to the position of the dynamical center. The shape of the pv-diagram
(Fig. \ref{abb07})
along the kinematic major axis is different from that of the strong
and extended \COe line emission (Fig. \ref{abb07}). The shape of the 
HCN(1-0) pv-diagram
is similar to the inner part of the \COz pv-diagram at
7 km/s (Fig. \ref{abb08}) convolved to a lower resolution.
This indicates that most of the HCN(1-0) emission is coming from a
region with $\leq$ 0.6'' diameter. 

\section{\label{ww12}MOLECULAR GAS KINEMATICS}

In this section we describe the properties of the molecular gas
kinematics in the circum-nuclear region of NGC~3227.
We outline how we derived a rotation curve and what the 
expected dynamical resonances in conjunction with the possible
kpc-scale bar in NGC~3227 are.

\subsection{\label{ww17}General kinematic properties of NGC~3227}

Mundell et al. (1995b) detected in their HI study of the system
NGC~3226/7 an HI cloud or
dwarf galaxy about 60'' (5 kpc) west of the nucleus of NGC~3227. 
The observed features (tidal tails, enhanced star formation at the
position of the western spiral arm) indicate that the interaction with
NGC~3226 and probably also the HI cloud is ongoing. On the other hand
the HI disk is relatively undisturbed, since its velocity field is in good
agreement with an inclined rotating disk (Mundell et al. 1995b).
\\
Gonz\'{a}lez Delgado \& Perez (1998) observed HII regions in
NGC~3227 in their H$\alpha$ emission. The HII regions show an offset
to the NW and SE relative 
to the major axis of the galaxy in agreement with the position of a
bar. Theoretical calculations (e.g. Athanassoula 1992a) predict that
gas at the leading side of the bar is shocked and compressed, favoring 
the formation of stars there. This indicates that
NGC~3227 is rotating clockwise. Therefore the spiral arms are trailing
and the southwestern side is
closer to the observer. The location of prominent dust lanes seen in
the optical HST images of Malkan et al. (1998) are consistent with
this geometry. 

\subsection{The nuclear molecular gas}

{\it The peculiar nuclear kinematics:}
In the \COz data set with a spectral resolution of 7 km/s the contrast of the 
more compact structures was enhanced over that of the more extended
components to allow for a detailed study
of the inner 1''. The pv-diagrams presented in
Fig. \ref{abb08} have now a nominal resolution of 0.3''. Along all 
position angles the emission in the inner 1'' does not drop linearly
to zero at the center. Along the kinematic major axis an apparent counter
rotation is observed between 0.2'' $\leq r \leq$ 0.5'' (see Fig. \ref{abb07}).
For even smaller radii the velocity flips back again to the rotation sense of
the outer structure at $r >$ 0.5''. This behavior forms a
{\sf S}-shape in the inner 1'' of the pv-diagrams. These changes in the 
rotation sense are present in all pv-diagrams. 
\\
\\
{\it The enclosed nuclear mass:}
If the velocity of the nuclear emission inside $r \leq$ 0.2'' is
due to Keplerian motion of molecular gas these data can be
used to estimate the enclosed mass.
Assuming an inclined disk the position angle 
under which these two emission regions have the largest angular separation 
of 0.28" at a position angle of PA $\sim$ 110$^o$. 
This is not coincident with the position angles of other
components like the radio jet ($\sim$ -10$^o$; Mundell et al. 1995b),
the [O~III] ionization cone ($\sim$ 15$^o$; Schmitt \& Kinney 1996) or
the H$\alpha$ outflow ($\sim$ 50$^o$; Arribas \& Mediavilla 1994).
Therefore a different cause for this high velocity except motion in a
nuclear potential appears to be unlikely.
The extent of 0.28'' at $PA$ 110$^o$ translates
into a radial distance of about 12~pc. Together with a velocity
difference  of
$\Delta v(12~pc) \sim 75 km/s$ (not corrected for inclination) this gives
a lower limit for the  enclosed mass of about 1.5 $\times$ 10$^7$ \solm. 
This limit is in approximate agreement with the enclosed mass 
derived for the central black hole in
NGC~3227 using H$\beta$ reverberation mapping 
with a BLR size scale of $\sim$ 17 days and a FWHM 
of the H$\beta$ line of 3900 km/s. 
With this technique Ho (1998) finds an enclosed mass of 
3.8 $\times$ 10$^7$ \solm ~whereas Salamanca et al. (1994) 
and Winge et al. (1995) estimate a black hole mass of $\sim$ 10$^8$ \solm.

\subsection{\label{ww13}The Rotation Curve}

We derived rotation curves from the \COe and \COz data
(using the routine 'ROTCUR' from GIPSY).
 This routine does not correct for the effect of beam smearing.
However, as the peaks in the \CO velocity fields occur at distances of
about 3.5'' and the smallest beam has a FWHM of 0.6'' this should 
not affect radii below 3.0'' and outside the central 1''. Using our \CO
measured values for the position of the dynamical center and a systemic
velocity of v$_{sys}$=1110$\pm$10~km/s, we obtain $i$ = (56$\pm$3)$^o$ and 
$PA$ = (160$\pm$2)$^o$, which are in excellent
agreement with the values of Mundell et al. (1995b) derived from HI data. 
\\
\\
To obtain a rotation curve ranging from the center to the outer HI disk
($\sim$ 100'') our \CO rotation curve was combined with the 
HI data (Mundell et al. 1995b) from the 
literature (Fig. \ref{abb11}). To analyze our data we use two rotation curves:
(1) For the {\it central 0.5''} we assumed Keplerian
rotation in agreement with the enclosed mass estimates in the inner 25~pc 
until the velocities of the Keplerian curve dropped well below the \CO 
velocities derived from the observations. (2) For 
the decomposition of the motion with the program 3DMod (see section \ref{ww24}) we 
extrapolated the rotation curve from its value at r=0.5'' to 0 km/s at r=0.0''.
Comparing this extrapolation of the rotation curve towards the center
to the observed data allows us to find non-circular velocities
associated with the nuclear region. 
This curve is referred to as our model rotation curve. Despite this difference
the two rotation curves are basically identical for radii $>$ 0.7''.
Our \CO data allows us to obtain a rotation curve for the range {\it
0.5'' $<$ r $\leq$ 5''}.
A rise of the rotation velocity till about 3'' to 4'' 
is observed.
No reliable rotation curve measurements are available between $r = 5''$
(end of the \CO rotation curve) and $r = 22''$  (innermost point of 
HI rotation curve not affected by beam smearing).
To connect the \CO curve with the HI curve we used 
a Keplerian velocity fall-off 
starting at $r = 5''$.
For radii {\it r $>$ 22''} the HI rotation curve was fully adopted. 
\\
\\
In order to test the deduced rotation curve as well as the values for
the inclination and position angle we looked for differences between the observed
\CO velocity field and a velocity field derived from the model rotation curve.
In general the residuals (Fig. \ref{abb12}) in the difference
field show characteristic patterns for mismatched parameters
(see van der Kruit \& Allen 1978). 
In our case 
they are less than about $\pm$20 km/s and indicate that most
of the line emission can be described by the derived rotation curve
satisfying the assumption of a simple rotating disk.
An exception are the molecular
bar and 
a region about 1'' south of the center that show
residuals of $\sim$ 35 km/s.

\subsection{\label{ww21}Dynamical Resonances}

It is possible to estimate the position of the dynamical resonances
from the rotation curve in conjunction with the bar length (Fig.
\ref{abb13}).
These resonances can be compared to the distribution of the molecular gas
and the HII regions. 

\subsubsection{\label{ww22}Theoretical positions of resonances}

NGC~3227 has probably a bar as indicated
by optical and NIR observations (Mulchaey, Regan, Kundu 1997, De
Robertis et al. 1998) with 6.7~kpc -
8.4~kpc radius and a position angle of about -20$^o$ relative to the major
kinematic axis. The presence of a bar is also supported by the locations
of the HII regions (Gonz\'{a}lez Delgado \& Perez 1998) that are
consistent with a bar for radii $\leq$ 50'' (size of their
field of view). 
\\
Under the assumption that the end of the bar is close to the
corotation (CR) this gives an angular pattern speed of $\Omega_p$ =
(32$\pm$5) km/s, since the HI rotation curve (Mundell et
al. 1995b) is relatively constant at these radii. The bar
would have an ILR at $r \sim$ 20'' (1.7~kpc). At
these distances one can find the SE-region of the \CO line emission
with two associated HII regions (Nr. 14 and 15, Gonz\'{a}lez Delgado \&
Perez 1998) as well as in northeast direction three more HII regions
(Nr. 18 - 20). These \CO emission line regions and the associated HII
regions are consistent with the presence of an ILR. 
\\
However,
the gas distribution and kinematics in the inner 40'' are not easily described
by structures at the position of
resonances due to the large-scale bar. Possible reasons for this are
that the gas flow is disturbed due to interaction with NGC~3226, the
HI cloud, or the HII regions. Another possibility is that
the gas dynamics are just beginning to be influenced by the
bar or that a stable semi equilibrium has not yet been reached. The analysis 
is  hampered by the poor knowledge of the
rotation curve for radii 5'' $< r <$ 22''.

\section{\label{ww24}DECOMPOSITION OF GAS MOTIONS IN NGC~3227}

We were successful in decomposing the molecular gas motions in NGC~3227 in their
circular and non-circular component using 3DMod with a model rotation
curve (see section \ref{ww12}). 
The results of the decomposition are essential since 
they high-light complex non-circular features in the velocity field
that will be modeled in section \ref{ww30}. 3DMod uses as an input the intensity
map, the rotation curve and a velocity dispersion distribution to
calculate the 3-dimensional spatial cubes of these properties. These
spatial 3-dimensional cubes are rotated according to the inclination
and position angle. Afterwards they are merged to a
3-dimensional $xyv$ cube which can now be compared directly to the 
measured data cube. 
A detailed description of the decomposition algorithm used in 3DMod is
given in Appendix~A. 

\subsection{The Decomposition of the interferometric \CO data}

The best decompositions were achieved using a systemic velocity of 
$v_{sys}$ = 1110~km/s in agreement with the derived  rotation curve 
and visual inspection of the nuclear pv-diagrams. 
For the velocity dispersion a value of 30 km/s combined with a thin disk
(0.2'' = 1 resolution element) gave the best fit to the data. 
The decomposition shows that observed higher velocity dispersions are due to a
spatial superposition of circular and non-circular components. 
The results are summarized in Table \ref{vv07}. 
\\
About 80 \% of the total line emission in the inner 8'' $\times$ 8''
of NGC~3227 is in excellent agreement with line emission from gas
in circular motion. 
However, the decomposition of the  \COz data reveals 3 distinct 
{\it components in non-circular motion:}
the nuclear region, the molecular bar, and a  $\le$0.6'' knot 1'' south. 
The {\it nuclear region} accounts for 3\% of the total flux in the inner 8''
$\times$ 8'' and allows the following 
possible interpretations:
{\bf (1)} The velocity field of this component
is in agreement with a counter rotating disk seen at
a position angle of (41$\pm$3) km/s with 
an inclination of $\sim$ 32$^o$
and an axial ratio of $\sim$ 0.85.
At this inclination the nuclear
disk is oriented orthogonal to the galaxy plane 
as $\mid 32^o \mid + \mid 56^o \mid \approx 90^o$.
This disk is also responsible for most of the HCN(1-0) line emission
(as in the case of NGC~1068 data ;Tacconi et al. 1997). 
{\bf (2)} A different possible cause for the complex nuclear velocity field
could be radial motions. However, such motions never change the 
direction of rotation on the major kinematic axis
and can therefore be ruled out. 
{\bf (3)} Further possibilities are motions in a
bar potential or the warping of the molecular gas. We explore these
issues further in section \ref{ww30}.
\\
The {\it molecular bar} is about 4'' longer in the western direction than along
its eastern extension and accounts for $\sim$17\% of the CO line flux
in the inner 25'' $\times$ 25''. 
The NS offset between the two sides of the bar might 
reflect its width. 
Its velocity is blue-shifted
(on both sides of the nucleus) relative 
to the circular velocity by about 50 km/s.

\section{\label{ww28} MASS AND THICKNESS OF THE GAS DISK}

A knowledge of the molecular gas mass and the dynamical mass of the
nuclear region is required in order to estimate the thickness of the
gas disk  as well as the torques acting on it in the 
case of a possible warping (see section \ref{ww30} and Appendix C).
A comparison of the molecular gas mass to the dynamical mass of the
nuclear region also shows that the molecular gas is a probe of 
the nuclear gravitational potential in
NGC~3227 rather than providing a dominant component of it.
\\
The molecular gas masses of the various component in the interferometric maps
are given in Table \ref{vv07}. 
We used the \nhico-conversion factor of 2 $\times$
10$^{20}$ $\frac{cm^{-2}}{K km/s}$ from Strong et al. (1989). 
(see e.g. Schinnerer, Eckart \& Tacconi 1998 for discussion
and references).
In addition we estimated the dynamical mass by using the inclination
corrected circular velocity for a given radius via
$M_{dyn} [M_{\odot}] = 232 \times v_{rot}(r)[km/s]^2 \times r[pc]~~.$ 

\noindent
The molecular gas contributes about 6 \% to the dynamical mass in
the inner 50~pc. 
We find an
average velocity dispersion $\sigma_{obs} \approx 30 km/s$ in the areas
of the dispersion map (2$^{nd}$-order moment) that show circular motions. The
velocity gradient of the rotation curve in the region from 0.35'' to
2.0'' is about 68 km/s arcsec$^{-1}$. This translates into an observed
velocity spread with a $FWHM_{rot}$ of
$\sim$ 41 km/s for a beam of 0.6''. The resulting velocity dispersion
is $\sigma_{rot} = \frac{FWHM_{rot}}{2 /sqrt{ln(2)}} \approx 25 km/s$.
We therefore find an intrinsic velocity dispersion $\sigma_{real} \approx 17
km/s$ in the inner 6'' of NGC~3227 using quadratic deconvolution.
This implies that the symmetrical structures in the
pv-diagrams that result in an increased apparent velocity dispersion have 
to be explained via a complex (ordered) velocity field  
rather than by an increased turbulence of the central molecular gas.
Following the equations used in
Schinnerer et al. (1999; see also
Quillen et al. 1992,
Combes \& Becquaert 1997,
Downes \& Solomon 1998)
we derive for the inner 550~pc a molecular gas disk height of about 15~pc.

\section{\label{ww30}RESULTS OF THE KINEMATIC MODELING}

To analyze the complex kinematics in the inner 50~pc of NGC~3227 showing
clear deviations from pure circular motions we
modeled the data with 3DRings (see appendix B). 
3DRings  allows to model non-circular
motions (1) via elliptical orbits with changing position angles
characterizing gas motions in a bar and (2) via circular orbits leaving
the plane of the galaxy representing a warp. 
The best bar solution fails to fully explain the data whereas the warp 
model gives a very satisfactory fit to the data.
\\
The model subdivides the disk into many individual (circular or
elliptical) orbits of molecular gas.
The inclination, position angle and shape of the
rotation curve for the overall galaxy were held fixed.
Each fitting process was started at large radii and successively
extended towards the center.
In  each case we tried several start set-ups that
all converged to similar (best) solutions with mean deviations 
from the data of less than about
10 km/s and 0.1'' for each radius and velocity in the pv-diagrams and
10$^o$ in the position angle of the mapped structures.
To test the quality and uniqueness of a model 
we used these criteria to derived the internal errors of the various model 
parameters 
($\alpha_0$, $\xi \Delta t$ , $\omega(r)$ for the warp
and $\epsilon(r)$, $PA(r)$ for the bar approach; see appendix B).
For both approaches we used a rotation curve in which the motions of
the inner few parsecs were assumed to be Keplerian due to the presence
of an enclosed mass. The best results for the
warp model is consistent with an enclosed 
mass of 2 $\times$ 10$^7$ \solm. 
\\
The pv-diagrams, the velocity field and the intensity map were
used as a guidance during the fitting process. 
To allow for an easy comparison between the data and the models, we
also displayed the pv-diagrams at an angular resolution of 0.3'' in order to
enhance in the cleaned data the contrast of the small scale
structures (Fig. \ref{abb08}).
All source components in this representation of the data can also
be identified in the
images at the nominal angular resolution of 0.6''.

\subsection{\label{ww31}The bar approach}

Elliptical orbits caused by a bar potential are a generally accepted way
to explain non-circular but well ordered motion in external galaxies
in the presences of a bar potential.
This is the only possibility to describe non-circular planar motion
that is stable for several orbital time scales. However, high angular
resolution NIR data shows no evidence for a strong nuclear bar 
(Schinnerer et al., in prep.).
In Fig. \ref{abb15} we show the curves for the position angle $PA$ and an
eccentricity $\epsilon$ that describe the ellipses for our best fit in
the framework of the bar approach.
Fig. \ref{abb16} shows that we are not able to account for the observed amount
of counter rotation along the kinematic major axis and especially along
PA~40$^o$ close to the kinematic minor axis. Also, in all pv-diagrams the 
{\sf S}-shape in the inner 1'' is not fully reproduced.
However, the model fails completely to reproduce the pv-diagrams
close to the kinematic minor axis, especially, the counter rotation
observed at $r \approx$ 0.5''. Also the fit can not explain the second
flip with the velocity rising according to the enclosed central
mass. The bar model can not fully reproduce the central 1'' 
intensity map and velocity field
(Fig. \ref{abb17})
for this bar model.
\\
Comparison to the calculated velocity fields and rotation curves of Wozniak 
\& Pfenniger (1997) who used self-consistent models of barred galaxies
shows that no strong apparent counter rotation along the minor axis is
possible in such a scenario (their figure 5). 
This makes us confident that our bar approach is valid and that the bar 
scenario is a less likely solution for the central 1'' in NGC~3227.

\subsection{\label{ww32}The warp approach}

Since the bar approach failed to fully explain the complex but well ordered
motion in the inner 1'' of NGC~3227
we inspected the second possibility to
explain non-circular motion: the warping of the gas disk mimicked by tilting
circular orbits out of the plane of the galaxy. 
This approach seems reasonable as gas can leave the plane of the
galaxy at the position of vertical resonances 
(Pfenniger 1984; Combes et al. 1990) and therefore probably
populate the essential orbits. As warps have also been observed in
accretion disks (e.g. NGC~4258 by Miyoshi et al. 1997) this seems a
plausible way to pursue. 
\\
In Fig. \ref{abb19} we
show the excellent fit of our warp model to the pv-diagrams. The intensity 
map and velocity field of this model are shown in Fig. \ref{abb20}.
The remaining differences 
(mainly reflecting the uneven intensity distribution in the data)
are mostly due to the assumption of a uniform
density distribution. The $\omega(r)$ curve (Fig. \ref{abb18})
required to fit the data is remarkably smooth.
The gas disk warps itself covering the nucleus starting from the south
at $r \sim$ 100~pc (1.2''). At $r \sim$ 30~pc (0.36'') 
the warped disk is orthogonal to the
host plane in agreement with the disk interpretation in 
section \ref{ww24}. At smaller radii the warp continues and the curvature
becomes stronger. In this geometry the AGN is obscured at least ones.
\\
The best fits result in $\alpha_o = (-120\pm20)^o$ with respect to the major
axis. The corresponding $\xi \Delta t$ is $(-2.7\pm0.7)\times10^5
yrs$. Therefore 
$\sim$10 rotations  at $r = 0.3''$ are required 
in order to precess by 360$^o$.
For  $\xi \Delta t > 0$ an $\alpha_o$ exist
such that the resulting solution is identical in its kinematics. This
indicates that there exists a single geometrical solution both with 
prograde and retrograde
precession relative to the rotation of the host galaxy.
To fit the counter rotation
it is required that  $\omega(r)$  rises above 90$^o$.
This is in agreement with Pringle (1997) who has shown that in the case of
accretion disks it is possible to obtain a stable warp till tilting 
angles of 180$^o$. 

\subsection{\label{ww33}Discussion of both approaches}

The comparison of both best solutions shows that the warp model is a
much better and preferred description of the data. The greatest
shortcoming of the planar bar model is that it fails to
reproduce the counter rotation along the minor axis, although the fit
is satisfying along the kinematic major axis. 
As a strong stellar bar is not observed in the NIR
(Chapman et al. (1999) and our SHARP K-band speckle image reconstructions)
strong streaming motions can not be evoked to explain the
remaining differences between the bar model and the data. These facts combined 
with the poor fit to the kinematic minor axis $pv$-diagram and to the 
observed spectra (Fig. \ref{abb22}) makes it not straight forward to 
explain the kinematics in the inner 0.8'' with a bar, i.e. via a pure planar 
system with ordered (elliptical) orbits.
Therefore our modeling suggest that the molecular nuclear gas disk in
NGC~3227 is likely to be warped in the
inner 70~pc. Also it is indicated that the gas observed at radii
$\geq$ 13~pc is relatively uniformly distributed rather than constrained
to some distinct areas in the nuclear region. 
\\
\\
The warp approach results in a very good fit to the $xyv$ data cube 
(see also spectra in Fig. \ref{abb22}) and therefore we definitely 
favor this model.
Theoretical analyses of the bulge formation and the bar dissolution using 
orbit calculations or N-body simulations show that with a sufficient
central mass density 3-dimensional stellar orbits form that can support
and enhance the so-called boxy or peanut appearance of bulges. 
The interesting zone is the region where the bar potential is only as 
strong as the bulge potential of the central mass. 
This happens often at about the distance of the radial ILR. 
Then in addition to the radial ILR a inner vertical resonance (IVR)
forms which allows stars to leave the plane of the disk (Pfenniger
1984, Combes et al. 1990). 
Friedli \& Benz (1993) have shown that even well within the plane of a 
barred galaxy vertical resonances can be high-lighted by the gas being 
pushed out of the plane. 
Since $x_4$ orbits are vertically unstable, the gas can leave the plane 
very quickly after formation of the bar and becomes trapped  in
stable anomalous orbits inclined with respect to the major axis $x$
(ANO$_x$ orbits, see Pfenniger \& Friedli 1991).
Similarly Garcia-Burillo et al. (1999)
observe in the warped galaxy NGC~4013 substantial amounts of molecular 
gas well above the plane.
In this case, however, the star formation activity in the disk might be
responsible. 
For NGC~3227 (and NGC~1068, Schinnerer et al. 1999) we find that 
the gas disk starts to warp at a radius at which the bulge begins to dominate 
the gravitational potential.  
Once out of the plane, the gas would be mainly 
supported by the potential of the bulge, competing with that of the bar or disk.
As the gas is dissipative compared to the stars it is valid to
assume that it will stay on ordered non-crossing orbits which will
strongly favor the formation of warps.
\\
\\
In order to move the gas out of the plane a torque is needed. Torques
can be induced by a non-spherical galactic potential (similar to the
effect of the halo on the HI disk), by the radiation pressure of the
radio jet (similar to the central radiation source causing the warp in
the accretion disk; Pringle 1996, 1997), by gas pressure in the
ionization cone (Quillen \& Bower 1999) or as a transient phenomenon
by the gravitational force of a dislocated molecular cloud complex
(see also appendix C for a more detailed discussion).
Estimates of these effects in the nuclear region of NGC~3227 
(see table in appendix) show
that torques induced by gas pressure or GMCs are the two most likely
causes. 
\\
\\
The only worrying aspect of the warp model is seen in the 3-dimensional view
(see Fig. \ref{abb21}). The direct view onto the AGN is seemingly blocked by
the warped gas disk. This is in contradiction to the unified scheme
for Seyfert galaxies that proposes for Seyfert~1 types a clear view
to the central engine. This problem can be softened,
if the orbits are not homogeneously filled with molecular gas but
rather with molecular clumps smaller than the size of the Seyfert~1
nucleus. In this case no effective shadowing of the compact Seyfert
nucleus itself can occur. 
\\
\\
The interpretation of
the X-ray data demands a warm absorber which would be located in the
immediate vicinity of the AGN (Komossa \& Fink 1997).
Recent observations of absorption lines in the UV of Seyfert~1 galaxies
show that these absorptions occur in galaxies with a warm absorber
(as NGC~3227; Crenshaw et al. 1999) at larger 
radial distances than the BLR.
\\
\\
The different classifications as a Seyfert~2 or Seyfert~1 found for
NGC~3227 in the literature are most likely caused by variability. 
NGC~3227 was first classified as a Seyfert~2 by Khachikian \& 
Weedman (1974). However, from data with higher S/N Osterbrock (1977)
identified it as a Seyfert 1.2. Further measurements by Heckman et
al. (1981) and Peterson et al. (1982) confirmed
this classification. On the other hand Schmidt \&
Miller (1985) note that the general low (and variable) strength of the
emission line spectrum is in better agreement with a Seyfert~2
nucleus. This indicates that the nucleus of NGC~3227 is either intrinsically
relatively weak for a Seyfert~1 galaxy or it is weakened by obscuration.
Obscuring gas in the warped molecular disk may well be responsible for
the observed variability  and varying classification of the nuclear source in 
NGC~3227.
\\
\\
It is also possible to compare the warp model with results from optical
polarimetry. Thompson et al. (1980) as well as Schmidt \& Miller
(1985) have observed a polarization of about $\sim$ 1\% for the
continuum as well as the permitted and forbidden emission lines
of the BLR and NLR. The
degree and position angle of the polarization for the continuum and
the emission lines are similarly implying a common cause for the
polarization. 
The position angle (131$^o$$\pm$8$^o$)
of the polarization is not in agreement with the galactic major axis
or the axis of the radio jet but it is in agreement with the
apparent enhancement of disk material due to projection effects 
in the warp geometry as proposed here.
\\
\\
{\it The described circumstances provide a strong independent support 
both for the presence of material
on the line sight to the nucleus as well as for the particular warp
geometry we obtained by analyzing the kinematics of the molecular gas.}

\section{\label{ww34}SUMMARY AND IMPLICATIONS}

\noindent
1. {\it Molecular gas close to the nucleus.---} We obtained PdBI data of 
the HCN(1-0) and the \CO line emission of the nuclear region in NGC~3227 with
sub-arcsecond spatial resolution. These data allow for the first time
a detailed and quantitative analysis of the molecular gas kinematics
in the inner
500~pc of this Seyfert~1 galaxy. NGC~3227 shows a nuclear gas ring with a
diameter of about 250~pc similar to the one observed in NGC~1068
(Tacconi et al. 1997, Schinnerer et al. 1999). 
\\
Gas emission at a radius of about 13~pc is detected in the \COz line
emission in NGC~3227. This emission shows a remarkable velocity offset
to the systemic velocity and allows for the first
time to use the molecular
line emission for an estimation of the enclosed mass in the
inner 25~pc of about $\sim$ 1.5$\times$10$^7$ \solm ~(not correcting
for inclination effects). This is in agreement with estimates from
other wavelength ranges.
\\
\\
2. {\it The HCN(1-0) line emission is very concentrated.---}
Comparison between the HCN(1-0) and the \CO data to single dish
observations suggest that the \CO line emission is distributed in a
disk of FWHM $\sim$ 25'' whereas the HCN(1-0) line emission is
concentrated on the nucleus. The direct comparison to our high
resolution interferometric \COz data furthermore suggests that the HCN(1-0)
is mainly arising from a region of size $\leq$ 0.6'' which shows
unusual kinematical behavior.
\\
\\
3. {\it The nuclear molecular gas disk is likely to be warped.---} 
To model the nuclear 
kinematics in NGC~3227 observed in the \COz
line emission we used a modified tilted ring model which is able to
describe gas motions in a thin ($\sim$17~pc), 
warp disk as well as in a bar potential.
Our modeling of the nuclear kinematics with 3DRings suggests that a
warped gas disk provides a better explanation to the observed gas
motions than motion evoked by a bar potential. The warp of the gas
disk starts at an outer radius of $\sim$ 75~pc and is perpendicular to
the outer disk of the host at $\sim$ 30~pc.
This warping indicates an obscuration of the nucleus at small
radii by a thin gas disk. This is in agreement with findings at other
wavelengths which suggest an obscuration at radii larger than the BLR,
including parts of the NLR.
The most likely cause for the warping of the gas disk is the gas
pressure in the ionized gas cones as traced by the NLR.
This pressure results in a torque onto the gas disk.
This mechanism was recently also discussed by Quillen \& Bower (1999) as
a possible cause for the warp in M~84. 
\\
\\
4. {\it Small molecular gas tori are not needed.---}
Our observations of NGC~1068 (Schinnerer et al. 1999) and NGC~3227
suggest that warps of the circumnuclear gas disk may be common. 
Even though nothing in our observations can rule out the existence
of small molecular tori with radii $\le$25~pc our finding 
implies that not under all circumstances
the postulated molecular gas torus of most unified schemes for
Seyfert galaxies are required to obscure the nuclei.
A somewhat peculiar distribution of molecular gas or dust in the host galaxy 
as proposed e.g. by Malkan et al. (1998) appears to be more likely.
However, the obscuring gas and dust may still be in well ordered motion.

\small
{\it Acknowledgments:}
IRAM is financed by INSU/CNRS (France), MPG (Germany) and IGN (Spain).
We thank the staff on Plateau de Bure for doing the observing, and the
staff at IRAM Grenoble for help with the data reduction,
especially D.\ Downes, R.\ Neri and J.\ Wink. For fruitful 
discussions we thank A.\ Baker, D.\ Downes,
P.\ Englmaier, J.\ Gallimore, O.\ Gerhard, R.\ Maiolino, A.\ Quillen,
N.\ Scoville and L.\  Sparke. We used 
the NASA/IPAC Extragalactic Database (NED) maintained by the
Jet Propulsion Laboratory, California Institute of Technology, under
contract with the National Aeronautics and Space Administration.

\newpage


\clearpage

\clearpage

\clearpage

\begin{figure}
\begin{center}
\psfig{file=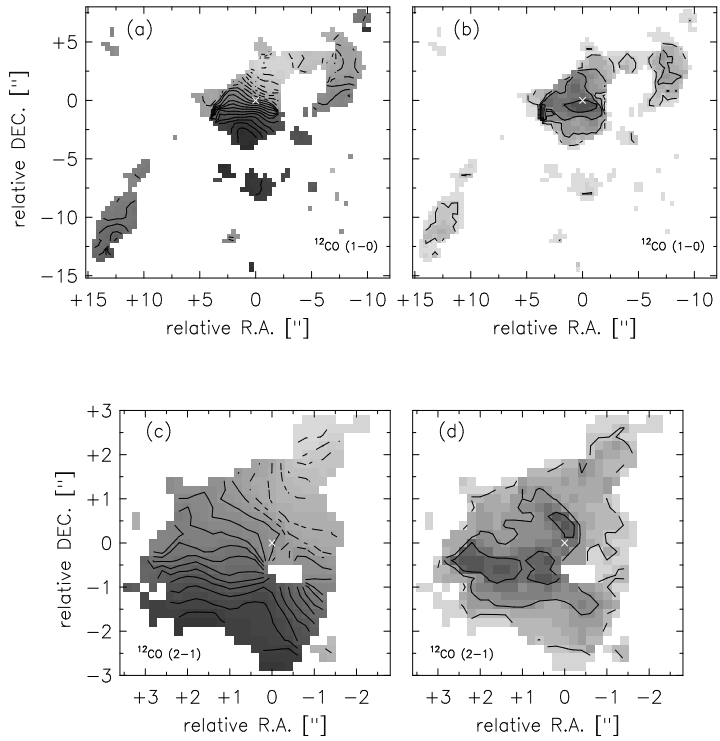,height=14.1cm,width=10.0cm,angle=0.0}
\end{center}
\figcaption[]{
\label{abb06}
Velocity field and map of the velocity dispersion in NGC~3227 as
obtained from the \COe line
emission ($a$ and $b$) and of the \COz line emission ($c$ and $d$).
The contours are at velocities of -220, -200, ... 160 km/s whereas the
first solid line corresponds to $\approx$ 1094 km/s
close to the systemic velocity of $v_{sys}$ = 1110~km/s.
The contours in the velocity dispersion maps are at 15, 30,
45 and 60 km/s, respectively.}
\end{figure}
\clearpage

\begin{figure}
\begin{center}
\psfig{file=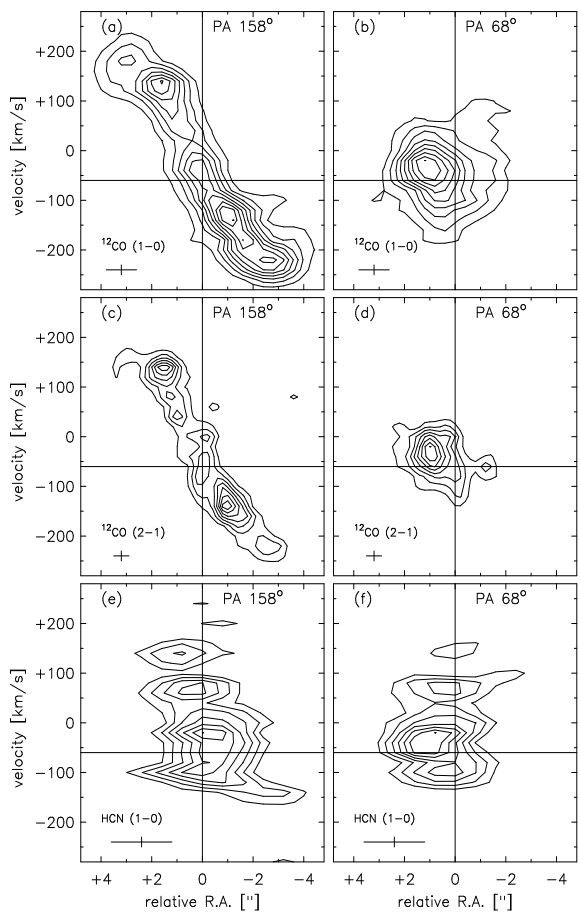,height=14.1cm,width=10.0cm,angle=0.0}
\end{center}
\figcaption[]{
\label{abb07}
pv-diagrams in NGC~3227 taken along the major and minor kinematic axis
of the \COe line emission ($a$ and $b$) and the \COz line emission ($c$
and $d$). The
contours are 20, 30, ... 100\% of the peak of the \COe emission and
30, 40, 100\% of the peak of the \COz emission.
pv-diagrams of the HCN(1-0) line emission ($e$ and $f$)
in NGC~3227 along the major and minor kinematic axis.
The contours are in 40, 50, ... 100\% of the peak
intensity.
}
\end{figure}
\clearpage

\begin{figure}
\begin{center}
\psfig{file=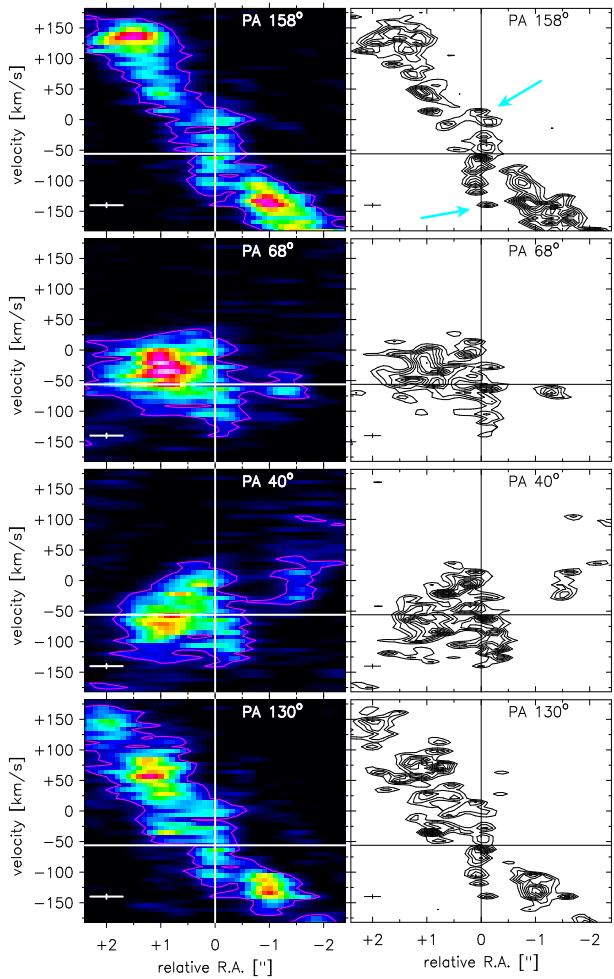,height=14.1cm,width=10.0cm,angle=0.0}
\end{center}
\figcaption[]{
\label{abb08}
pv-diagrams of the nuclear \COz line emission in NGC~3227 
along different position angles. 
In all pv-diagrams the fast change of the
velocity inside the inner 1'' is clearly visible as an {\sf S}-shape.
To highlight this complex velocity behavior
the data is also shown at a nominal resolution of 0.3'' (right).
For comparision the corresponding pv-diagrams at the achieved 
instrumental resolution of 0.6'' are shown in color on the left panels.
The (magenta) contour corresponds to 3$\sigma$, with 1$\sigma$= 6.2
mJy/beam at a spectral resolution of 7 km/s.
For the first time seen in the molecular gas emission the data show
a rising rotation curve towards smaller radii with the extreme
velocities (indicated by arrows) at radial separations of 
$\approx$0.15'' (13~pc) indicating an enclosed mass of 
$\ge$2$\times$10$^7$\solm, not correcting
for inclination effects.
This value is consistent with $\sim$10$^8$\solm ~estimated from
H$\beta$-reverberations mapping by Salamanca et al. (1994).
}
\end{figure}
\clearpage

\begin{figure}
\begin{center}
\psfig{file=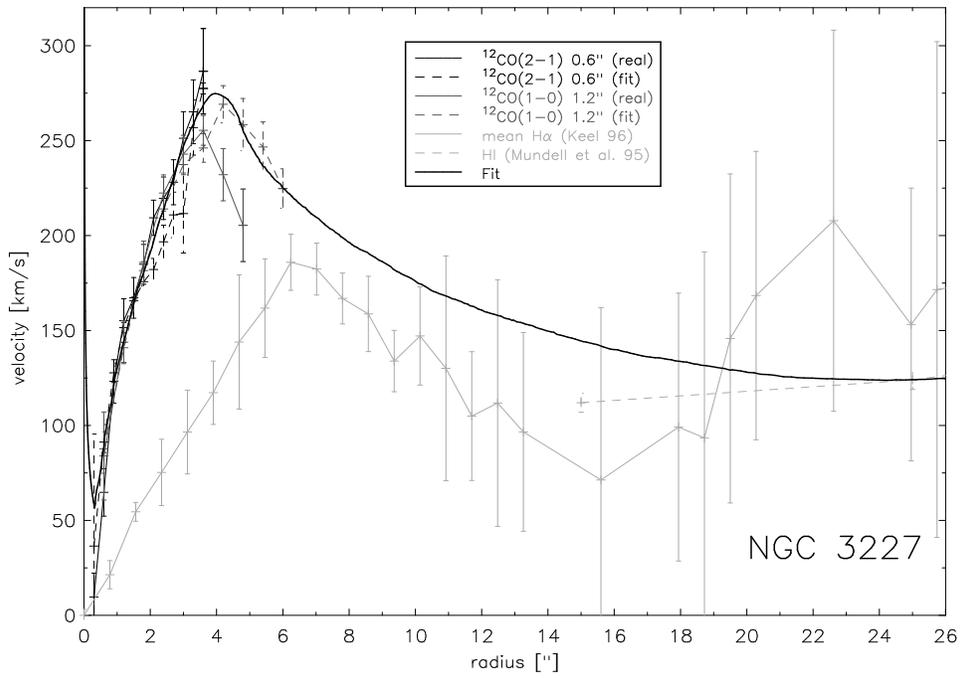,height=10.0cm,width=14.1cm,angle=-90.0}
\end{center}
\figcaption[]{
\label{abb11}
The rotation curves of NGC~3227. The rotation curves derived from the
\CO data as well as curves taken from the literature are shown. The
thick line is the best fit to the different curves. 
}
\end{figure}
\clearpage

\begin{figure}
\begin{center}
\psfig{file=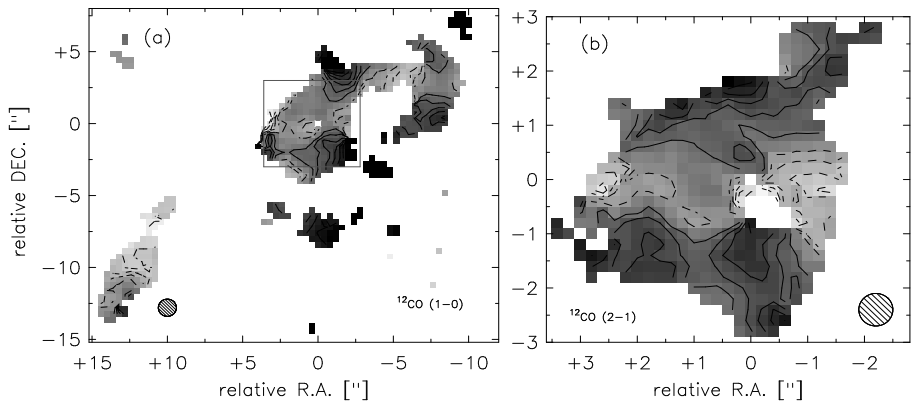,height=10.0cm,width=14.1cm,angle=-90.0}
\end{center}
\figcaption[]{
\label{abb12}
Residuals of the fit to the velocity field in NGC~3227 of the \COe
line emission (left) and the \COz line emission (right). Positive residuals are
shown in solid contours starting at 10 km/s in steps of 10 km/s,
negative residuals start at -10 km/s.}
\end{figure}
\clearpage

\begin{figure}
\begin{center}
\psfig{file=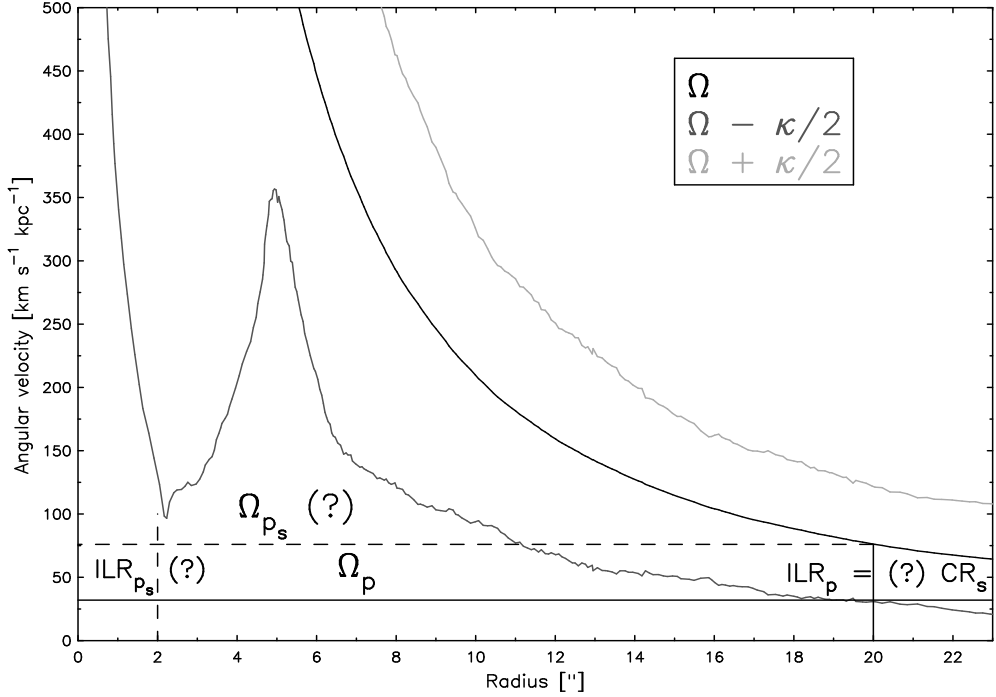,height=10.0cm,width=14.1cm,angle=-90.0}
\end{center}
\figcaption[]{
\label{abb13}
Dynamical resonances in NGC~3227. The angular velocity $\Omega$ as
well as the $\Omega + \frac{\kappa}{2}$ and $\Omega - \frac{\kappa}{2}$ 
curves are shown. The (primary) outer bar has an angular pattern
velocity $\Omega_{p_p}$ indicating an ILR$_p$ at about r=20''. The
resulting angular velocity for a possible (secondary) inner bar
$\Omega_{p_s}$ as well as its related ILR$_s$ are uncertain and
therefore labeled with a question mark. For further explanations see text.}
\end{figure}
\clearpage

\begin{figure}
\begin{center}
\psfig{file=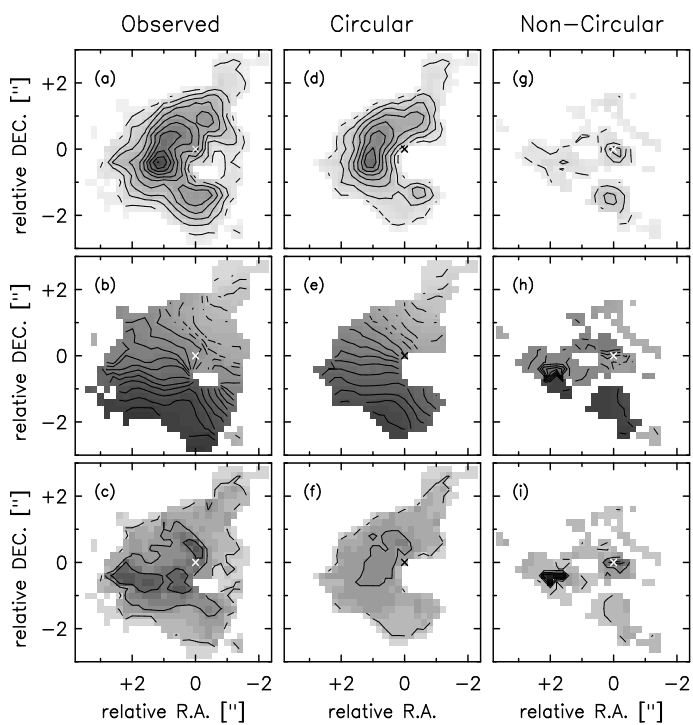,height=14.1cm,width=10.0cm,angle=0.0}
\end{center}
\figcaption[]{
\label{abb14}
Decomposition of the \COz data ($a$ - $c$) into components that follow
pure circular motion ($d$ - $f$) and those that follow  non-circular 
motion ($g$ - $i$) in NGC~3227.
The contours of the intensity maps ($a$, $d$, $g$) are in 10, 20, ...
100\% of the peak intensity in the observed map. In the velocity
fields ($b$, $e$, $h$) the contours are shown from -220 km/s till -80
km/s (broken line) and from  -60 km/s till 160 km/s (solid line) in
steps of 20 km/s. The contours of the velocity dispersion maps 
($c$, $f$, $i$) are at 15, 30, 45 and 60 km/s.}
\end{figure}
\clearpage

\begin{figure}
\begin{center}
\psfig{file=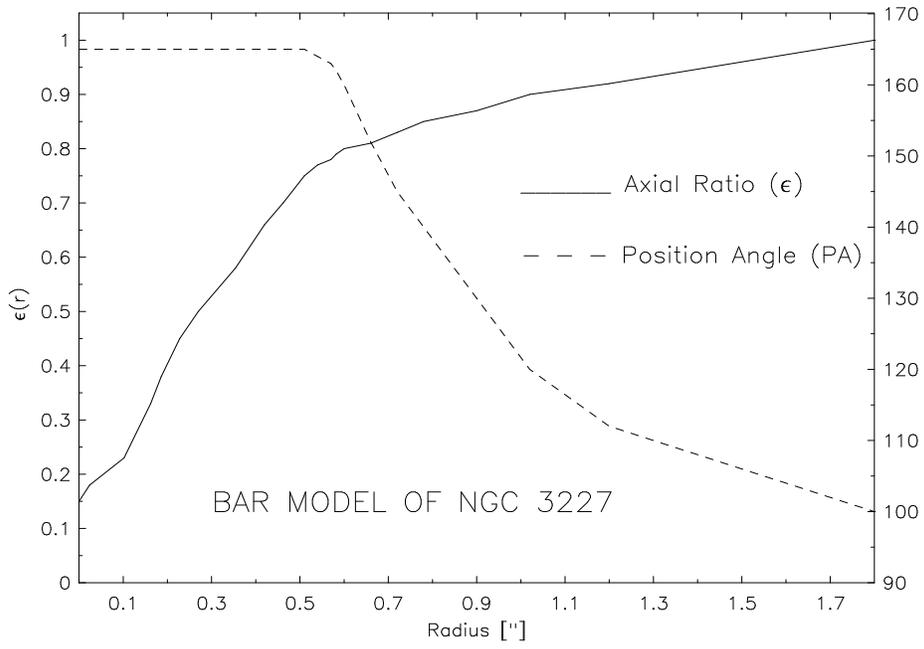,height=10.0cm,width=14.1cm,angle=-90.0}
\end{center}
\figcaption[]{
\label{abb15}
The curves of the position angle $PA$ and eccentricity $\epsilon$ 
of the bar approach in NGC~3227. The
$\epsilon$ curve (solid line) and the $PA$ curve (broken line) for the
best bar model. }
\end{figure}
\clearpage

\begin{figure}
\begin{center}
\psfig{file=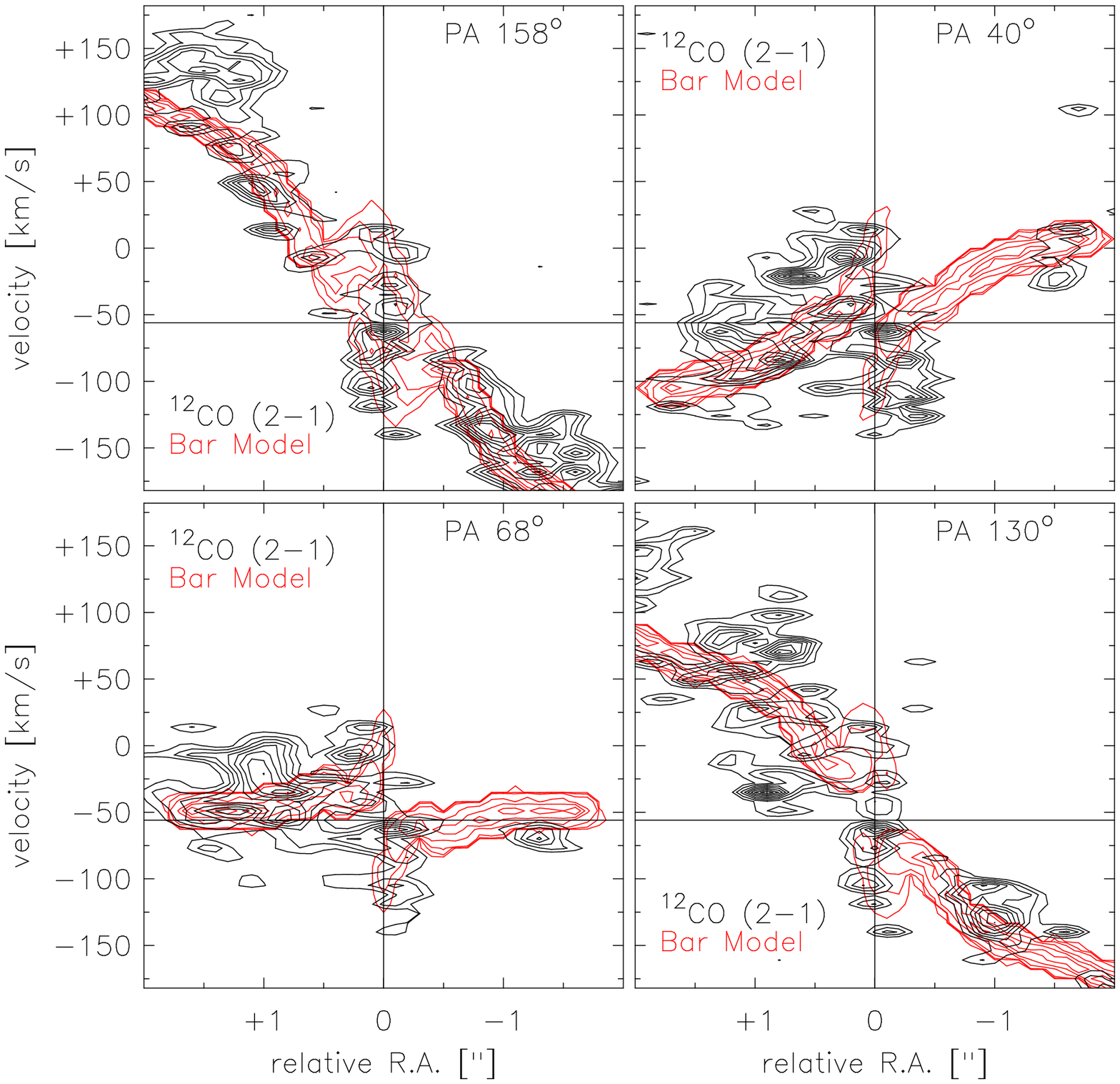,height=14.1cm,width=10.0cm,angle=0.0}
\end{center}
\figcaption[]{
\label{abb16}
pv-diagrams for the bar approach in NGC~3227.
The data (black contours) along different position angles
are shown together with the results (red contours) of the
bar model.
To highlight the complex velocity behavior
the data is shown at a nominal resolution of 0.3''.
In the pv-diagram at a resolution of 0.6''
the lowest contour corresponds to 3$\sigma$, with 1$\sigma$= 6.2
mJy/beam at a spectral resolution of 7 km/s.
}
\end{figure}
\clearpage

\begin{figure}
\begin{center}
\psfig{file=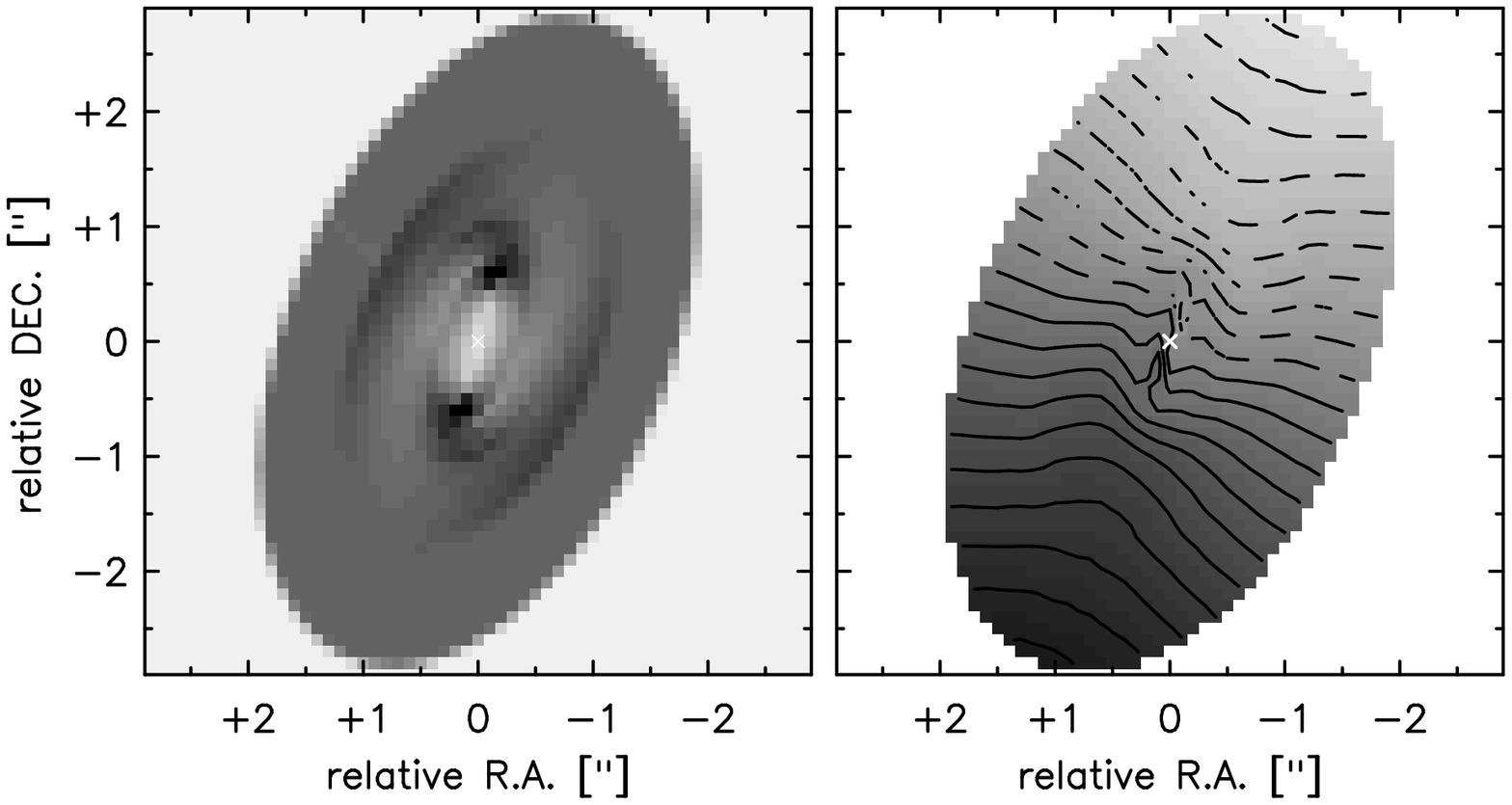,height=10.0cm,width=14.1cm,angle=-90.0}
\end{center}
\figcaption[]{
\label{abb17}
Intensity map (left) and velocity field (right) of the bar approach in
NGC~3227. The contours of the velocity field range from -200 km/s
to 200 km/s in steps of 20 km/s. The negative velocities are indicated
by broken lines.}
\end{figure}
\clearpage

\begin{figure}
\begin{center}
\psfig{file=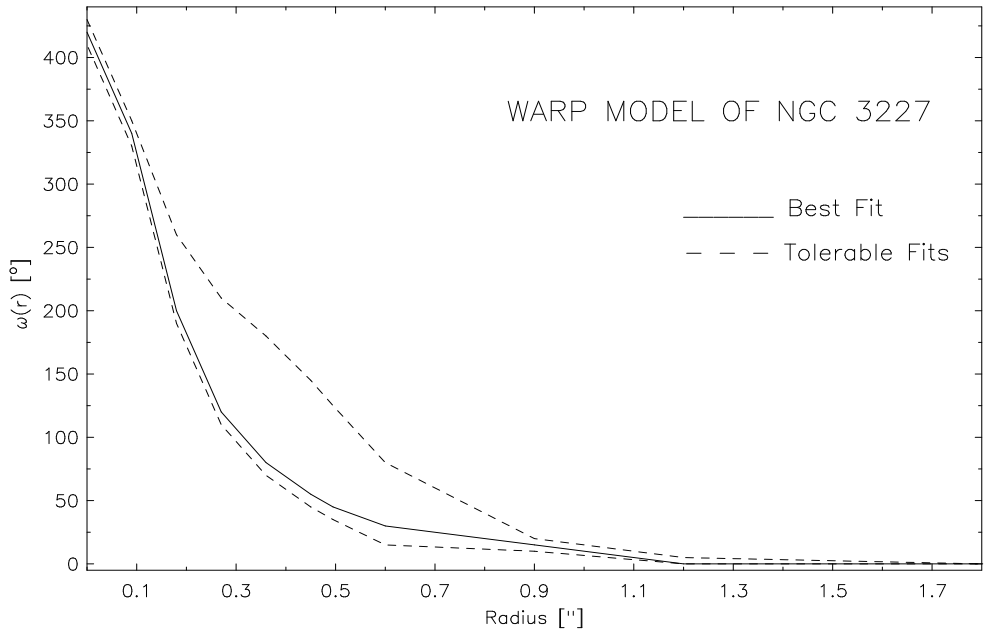,height=10.0cm,width=14.1cm,angle=-90.0}
\end{center}
\figcaption[]{
\label{abb18}
The $\omega(r)$ curve of the warp approach in NGC~3227. The solid line represents
the best fit to the data. The broken lines indicate the range over
which satisfying fits to the data are still possible.}
\end{figure}
\clearpage

\begin{figure}
\begin{center}
\psfig{file=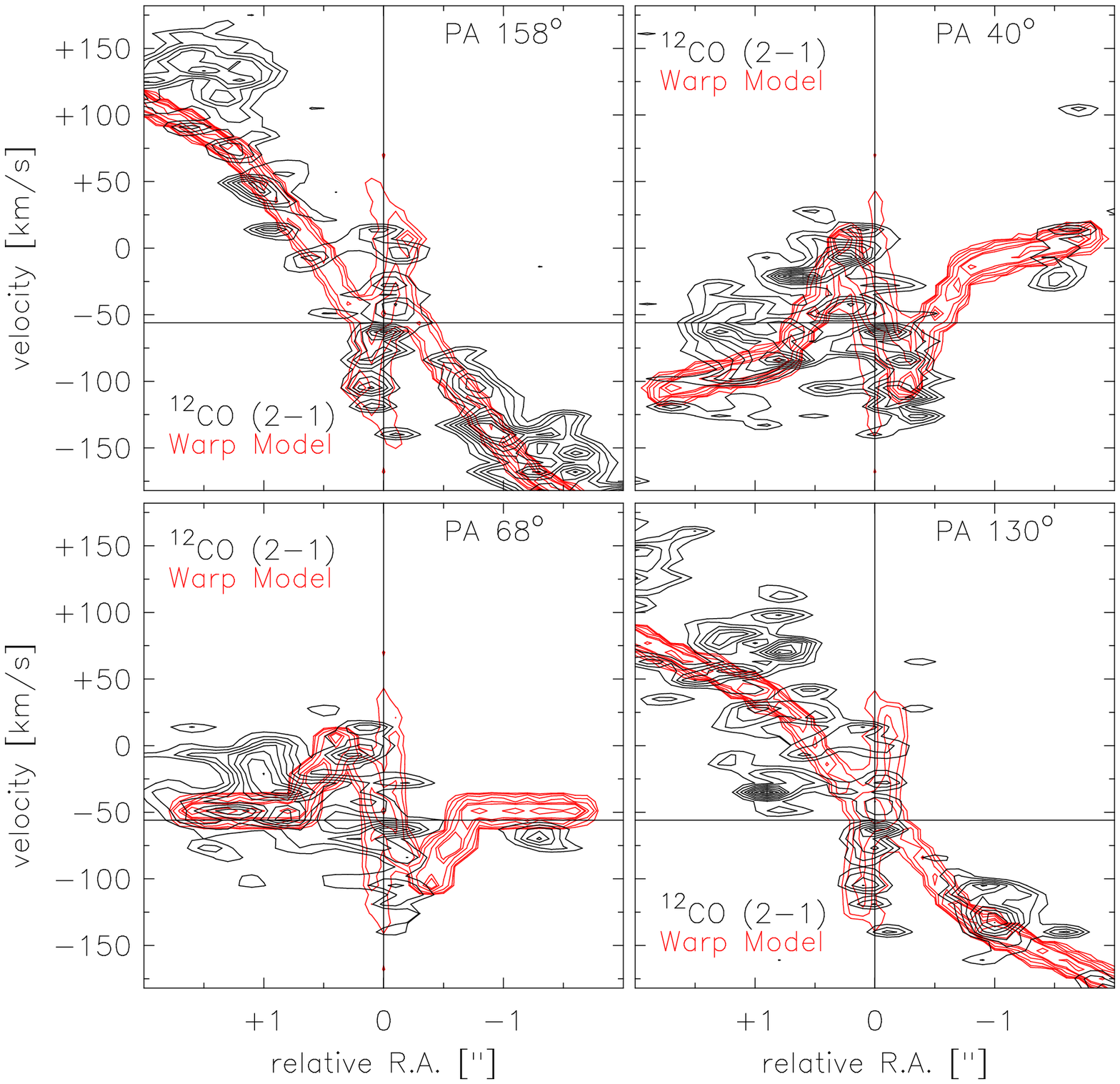,height=14.1cm,width=10.0cm,angle=0.0}
\end{center}
\figcaption[]{
\label{abb19}
pv-diagrams for the warp approach in NGC~3227.
The data (black contours) along different position angles
are shown together with the results (red contours) of 
the warp model.
To highlight the complex velocity behavior
the data is shown at a nominal resolution of 0.3''.
In the pv-diagram at a resolution of 0.6''
the lowest contour corresponds to 3$\sigma$, with 1$\sigma$= 6.2
mJy/beam at a spectral resolution of 7 km/s.
}
\end{figure}
\clearpage

\begin{figure}
\begin{center}
\psfig{file=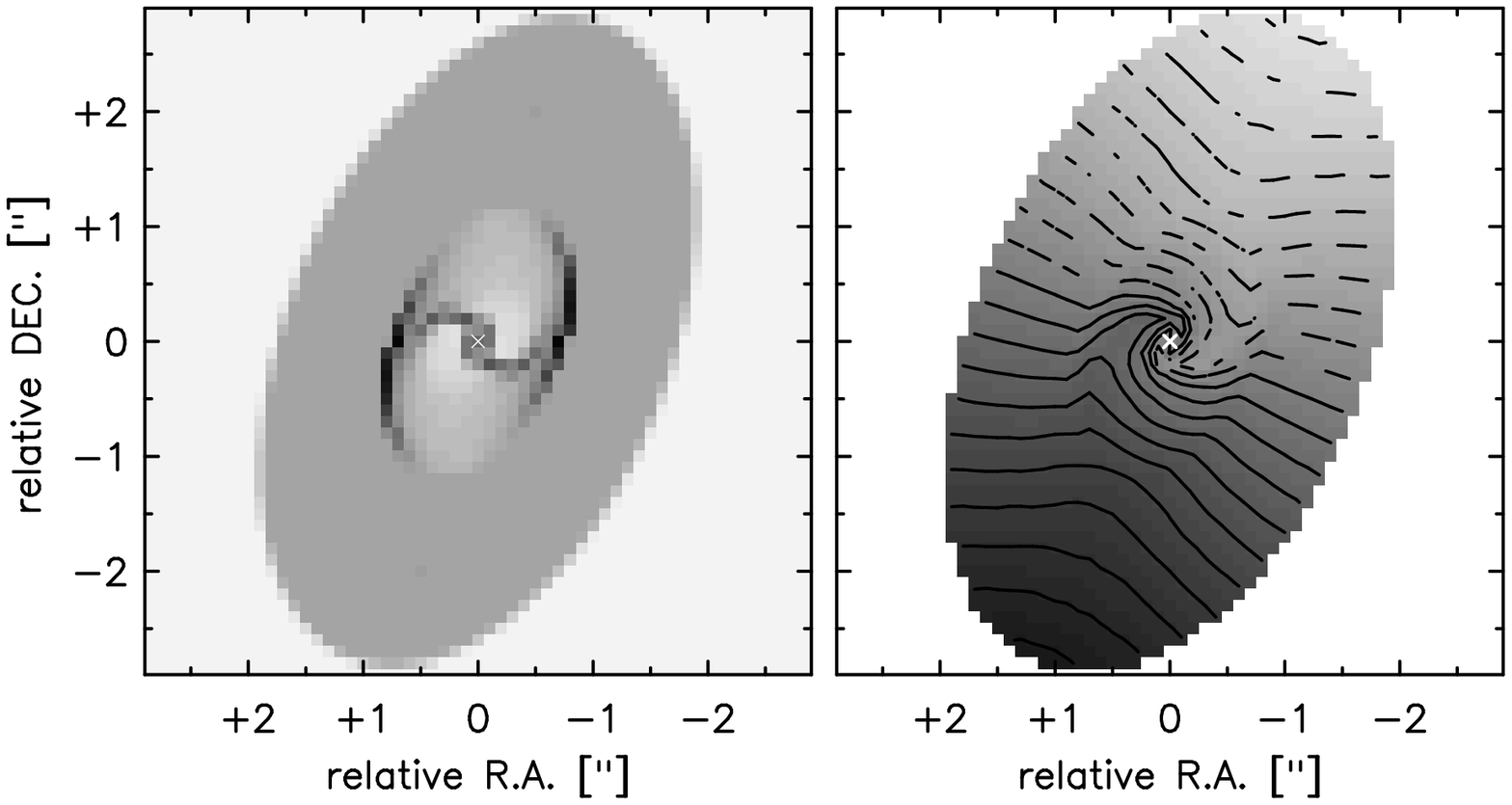,height=10.0cm,width=14.1cm,angle=-90.0}
\end{center}
\figcaption[]{
\label{abb20}
Intensity map (left) and velocity field (right) of the warp approach
in NGC~3227. The contours of the velocity field range from -200 km/s
to 200 km/s in steps of 20 km/s. The negative velocities are indicated
by broken lines.}
\end{figure}
\clearpage

\begin{figure}
\begin{center}
\psfig{file=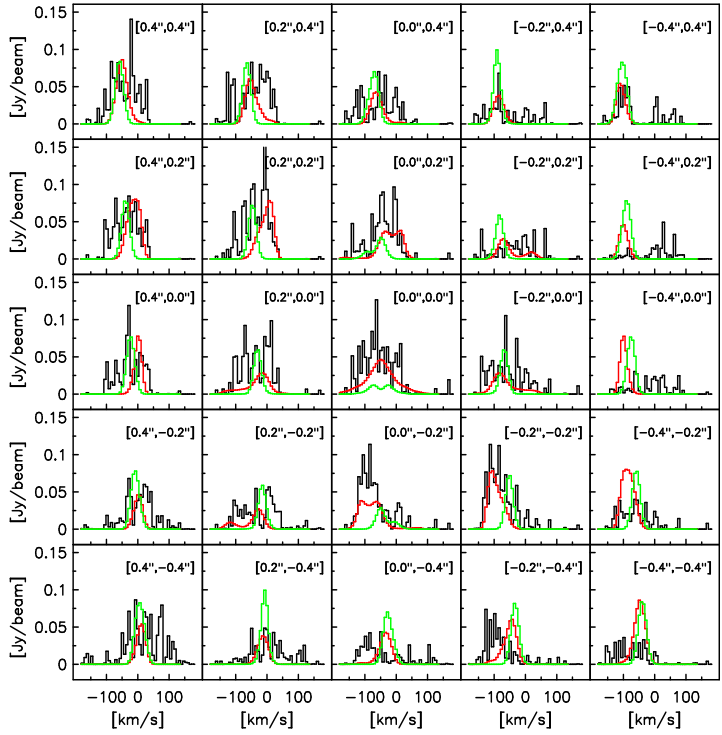,height=14.1cm,width=10.0cm,angle=0.0}
\end{center}
\figcaption[]{
\label{abb22}
Here we show a comparison of the measured spectra (black) and the spectra
given by the warp (red) and bar model (green). 
We concentrate on the
central 0.8'' $\times$ 0.8'' in steps of 0.2''. 
The warp model represents an acceptable fit to
the observed data that reflect the complex velocity field. The
agreement 
of the bar model with the data is not
as good as it cannot reproduce the asymmetric line shapes 
especially at positions north-east and south-west of the nucleus.
}
\end{figure}
\clearpage

\begin{figure}
\begin{center}
\psfig{file=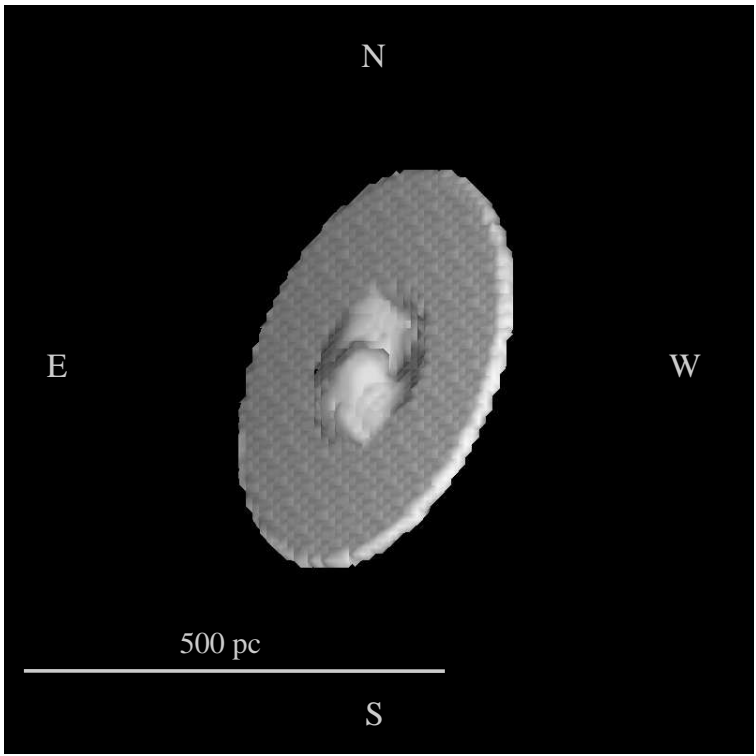,height=10.0cm,width=10.0cm,angle=0.0}
\end{center}
\figcaption[]{Spatial geometry of the warp model of NGC~3227. Bright
\label{abb21}
colors of the warped disk indicate sections that are closer to the
observer, darker colors those further away.
}
\end{figure}
\clearpage

\clearpage

\begin{table}[htb]
\caption{
\label{ww1}
Properties of NGC 3227}
\begin{center}
\begin{tabular}{lc}\hline \hline
 & NGC~3227  \\ \hline 
Right ascension (J2000) & 10$^h$ 23$^m$ 30.6$^s$ \\
Declination (J2000) & 19$^o$ 51' 53.99'' \\
Classification & SAB(s) pec \\
Inclination & 56$^o$ \\
Position angle & 158$^o$ \\
AGN type & Sey~1.2 \\
Systemic velocity v$_{sys; HI}$& 1135~km/s \\
Distance & 17.3 Mpc \\
1'' equals & 84 pc \\
\hline \hline
\end{tabular}
\end{center}
The sky coordinates, the systemic velocity as well as the Seyfert type
were taken from NED (NASA/IPAC Extragalactic Database). 
The classification is from the RC3 catalog (de Vaucouleurs et al. 1991).
Inclination and position angle are taken from Mundell et al. (1992b).
For the distance we adopted the value of the LGG 193 group (Garcia
1993) were NGC~3227 is a member.
\end{table}

\begin{table}[htb]
\caption{
\label{vv07} 
Components of the $^{12}$CO line emission in NGC~3227}
\begin{center}
\begin{tabular}{lrrrrrrr}\hline \hline
Component & R.A. & DEC.& $S_{CO} \Delta v$ & $\frac{CO(2-1)}{CO(1-0)}$ & M$_{H_2}$ & M$_{dyn}$ \\
                  &[''] & [''] & [Jy km/s] &  & [10$^{7}$ M$_{\odot}$] & [10$^8$ M$_{\odot}$] \\ \hline
nuclear region         &  0.0 &   0.0 & 3.8    & 0.7  & 0.19  & 0.34 \\
ring                   &  0.0 &   0.0 & 55     & 0.4  & 8.49  & 44.0 \\
red knot               &  0.0 &   1.0 & 4.9    & 0.5  & 0.37  &      \\
molecular bar          &  6.0 &   0.1 & 6.2    & 0.6  & 0.95  &      \\
NW-region              & -8.0 &  +3.0 & 11     & 0.3  & 1.70  &      \\
SE-region              &+12.0 & +10.0 & 6.8    & --   & 1.06  &      \\
S-region               &  0.0 &  -5.0 &  1.2   & 1.3  & 0.18  &      \\ \hline
8.4'' $\times$ 8.4''   &  0.0 &   0.0 & 75     & 0.5  & 11.60 & 44.0 \\
24.4'' $\times$ 24.4'' &  0.0 &   0.0 &  98    & --   & 15.20 &      \\
\hline \hline
\end{tabular}
\end{center}
Listed are the centroid positions of individual source components,
as separated by modeling (with 3DMod).
The intensities of the \CO lines given here are not corrected for the 
different beam sizes. 
However, this correction was taken into account for the line ratio. 
The intensity of the ring was measured in a disk with $r$=1.5'' in
the circular intensity map of gas that is in a disk with pure 
circular rotation (see Fig. \ref{abb14}). 
With the possible exception of the NW-, and SE-regions all 
other source components are not part of a simple circular rotating disk.
The corresponding \htwo masses of the individual components are 
derived in section \ref{ww24}.
The estimated errors of the integrated line intensity values are 15\% 
for the \COe line and 30\% for the \COz line.
For each component the flux and the obtained molecular \htwo mass 
M$_{H_2}$ is given. In addition we give the
dynamical mass M$_{dyn}$ if possible. 
\end{table}

\clearpage
\appendix

\large
\begin{center}
{\bf APPENDIX}
\end{center}
\normalsize

Here we give a detailed description of the algorithms we used to analyze
(Appendix A) and describe (Appendix B) the kinematics and intensity
distribution of the molecular gas in the nuclear region of NGC~3227.
We also summarize and compare possible physical mechanisms that may lead to a
warp of the molecular disk in the circum-nuclear region of active
galactic nuclei (Appendix C).

\section{DECOMPOSITION OF MOTIONS: 3DMod}

The application of 3DMod allows us to decompose the observed kinematics
into their components of non-circular motion and circular  rotation
around the nuclear position. 
As an input it uses information on the integrated line flux
distribution, the rotation curve and the spatial distribution of the velocity
dispersion.
The 3-dimensional spatial distributions of the intensity, the velocity and the
velocity dispersion are generated separately.
\\
{\bf Intensity distribution:} 
We use the deprojected measured intensity
distribution. We correct for resolution dependent deprojection effects by
deconvolution before the deprojection and a later reconvolution.
The de-projected deconvolved intensity map is loaded into the $xyz$
model cube also allowing to introduce a thickness of the molecular disk
in the $z$ direction, if required.
\\
{\bf The velocity field:}
The spatial velocity cube is constructed assuming that  the gas
is dynamically coupled via interaction between the
individual molecular clouds and clumps and
that the velocity field does not vary significantly with height $z$.
The rotation curve is  extrapolated to $v~=~0 km/s$ towards 
the origin (r~=~0) and used to
model the axisymmetric velocity field in the plane of the galaxy
assuming circular rotation.
\\
{\bf Velocity Dispersion:}
We assume that the velocity dispersion of the gas is locally isotropic.
We allow for a possible exponential variation
$\sigma(x,y,z) = A \times r(x,y,z)^{-\alpha}$.
Here $r(x,y,z)$ is the radial distance from the origin. Such a description is
appropriate for the transition between a possibly large nuclear velocity
dispersions and the low dispersions in the disk. A minimum velocity
dispersion of $7 km/s$ is assumed for the dispersion between the
individual clouds (e.g. Combes \& Becquaert 1997).
\\
{\bf Creation of Model Cubes:}
Each spatial cube is corrected for inclination $i$ and position angle
$PA$ of the galaxy.
After rotation the 3 individual spatial cubes are merged
to obtain a $xyv$ cube with two spatial axes 
(on the sky) and a spectral axis. The spectral with 
a resolution similar to the observed data is
generated via integration along the line of sight.
For each spatial pixel
in the plane of the sky the central velocity is determined and the
flux is distributed into the spectral pixels according the local velocity
dispersion. 
Convolution along the remaining two axis gives the
required spatial resolution.
After scaling to the observed flux 
the calibrated model $xyv$ cube can directly be compared to the
measured data.
\\
{\bf Decomposition of Motions:}
The $xyv$ cube is generated as described above 
assuming that all the flux is originating from gas in 
circular motion. The difference between the data cube and the model
cube then shows positive and negative residuals at all positions for which the
assumption of circular motion is not correct. In general it can be
demanded that the integrated flux at each position is positive. This
means that the positions which have negative residuals have to be
corrected. As a first approximation of the correction we create an intensity
map of the
negative residuals by integrating them over velocity, subtracting a
mean negative noise level, and then setting all positive values in the
resulting map to zero. This map is added to the observed intensity
map. The new map should now contain only emission from components in
circular motion and is used as a new input to calculate the intensity cube.
Then a new $xyv$ cube is generated and subtracted from the data cube.
The residual cube should now contain only significant positive residuals which
represent components in non-circular motion.

\newpage
\section{KINEMATIC MODELING: 3DRings}

\begin{figure}
\begin{center}
\psfig{file=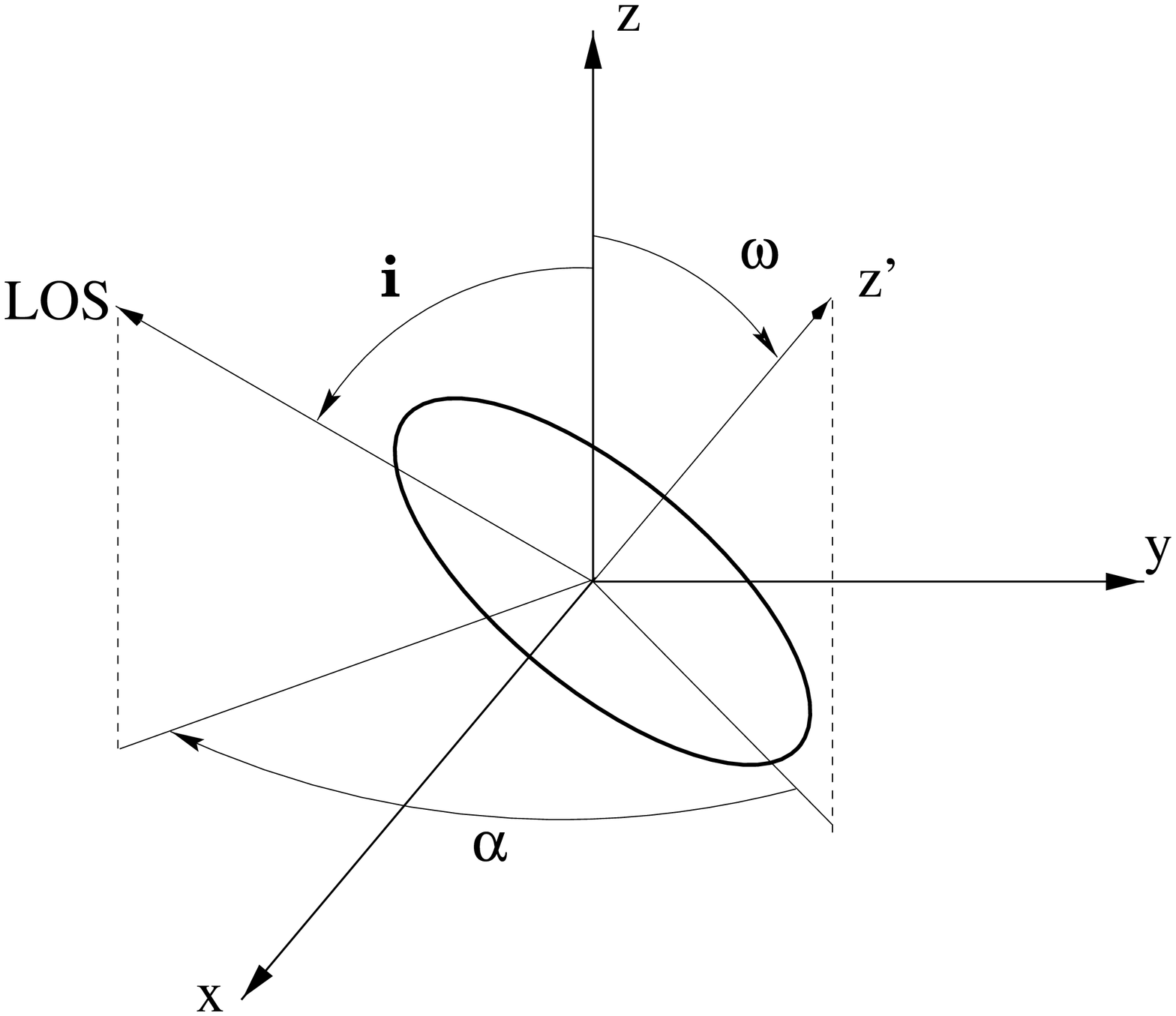,height=10.0cm,width=14.1cm,angle=-90.0}
\end{center}
\figcaption[]{
\label{abbb3}
Important quantities for the warp approach of the model 3DRings. A single
ring is shown which is tilted by an angle $\omega$ relative to the 
rotation axis of the galaxy which is inclined to the line of sight by the
angle $i$. Here $\alpha$ is the precession angle between the ring
axis and the projection of the line of sight onto the plane of the galaxy.}
\end{figure}

\begin{figure}
\begin{center}
\psfig{file=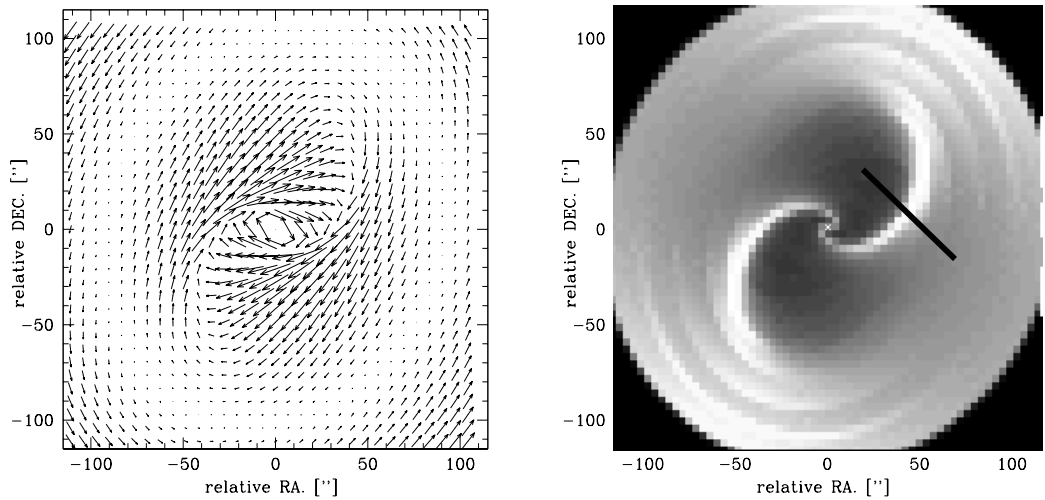,height=7.0cm,width=14.1cm,angle=-90.0}
\end{center}
\figcaption[]{
\label{abbb5}
Density distribution (right) and the velocity field (left)  
obtained from our model bar approach (from 3DRings). 
These results are comparable
to {\em Model 001} from Athanassoula (1992b). As in
Athanassoula (1992b) we subtracted from the total velocity field the
field of a rigid rotating disk with the angular velocity $\Omega_P$ at
the CO-rotation. In the density distributions (right) the slit positions
for Fig. \ref{abbb6} are indicated by a thick line.}
\end{figure}

\begin{figure}
\begin{center}
\psfig{file=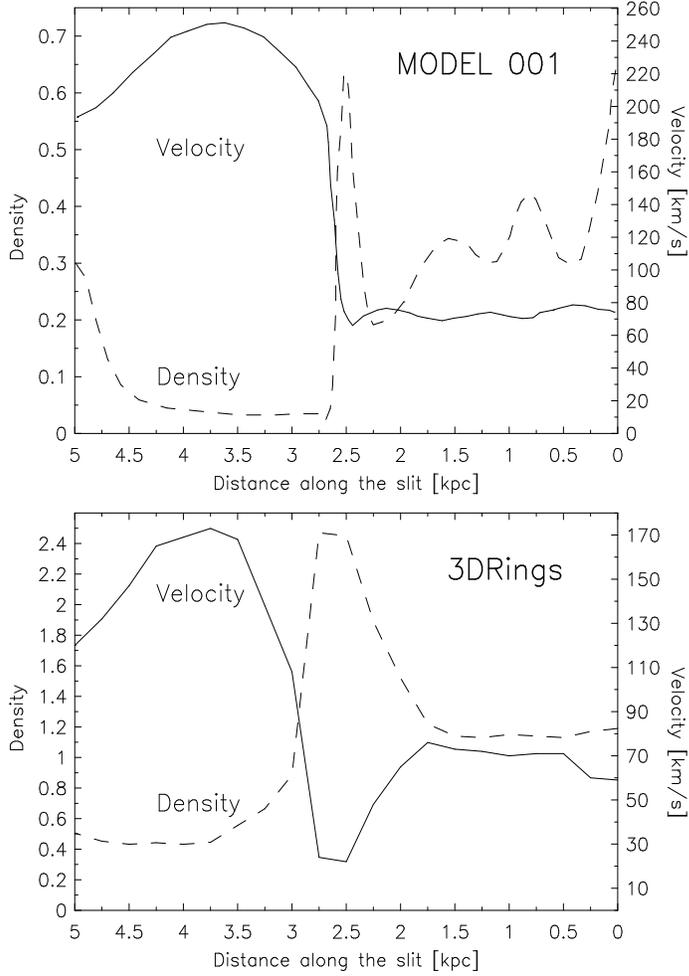,height=14.1cm,width=10.0cm,angle=0.0}
\end{center}
\figcaption[]{
\label{abbb6}
Comparison of the shock fronts between the {\em Model 001} from Athanassoula
(1992b) and our model calculations. Along the slit across the shock front
(indicated in Fig. \ref{abbb5}) the velocities (solid line) as well as density 
distributions (broken line) in {\em Model 001} (top) are shown for 
from SW (left) to NE (right). 
In both cases the abrupt
increase of the density as well as the strong drop in the velocity is
seen at the position of the shock front. This demonstrates that the bar
approach of 3DRings provides qualitatively (and within tolerable limits
also quantitatively) similar results as more complex theoretical
calculations. Thus 3DRings can be used to search for bars via the
fitting of the 3 dimensional data cube ($xyv$ cube).}
\end{figure}

The model 3DRings allows to create kinematic models of circular and
non-circular motions. 
The model is based on the following considerations: Since gas is
dissipative it can only move for a longer time on orbits which are not
self-intersecting, not crossing and do not have strong cusps.
Therefore only two basic possibilities are available for gas orbits: (1) planar
circular or elliptical orbits and (2) tilted circular orbits. The first
possibility can be identified with the $x_1$ and $x_2$ orbits 
(e.g. Contopoulos \& Papayannopoulos 1980)  which
exist in bar potentials in the plane of the host galaxy. The second
possibility describes gas motion that is no longer confined to the
plane of the galaxy so that the gas disk is warped.
We use the inclination, position angle, and the rotation curve 
derived from the molecular gas velocity field. 
The gas disk is subdivided into a number of elliptical
(bar approach) or circular (warp approach) rings between a given inner 
and outer radius. For both approaches the properties of
these 3-dimensional rings are given by continuous parameter input curves as
a function of radius.
The fitting is started at large radii where simple
circular motion dominates and successively extended towards the nuclear region. 
The fit is done matching the pv-diagrams and the moment maps. 
\\
{\bf The bar approach:}
Rather than calculating the gas motion in a given gravitational,
potential 3DRings fits the observed $xyv$ data cube
under the simplifying assumption of closed elliptical $x_1$ and $x_2$ orbits
with continous curves of ellipticity $\epsilon(r)$ and position angle $PA(r)$
and centered on the nucleus.
This approach has been adopted from Telesco \& Decher (1988; Fig.~7 therein)
who discussed its validity in great detail. 
To match observations and theoretical 
calculations we implemented the orbits such that
the velocities at the minor and major axis are inversely
proportional to the axial ratio.
\\
{\bf The warp approach:}
3DRings is similar to other tilted ring models (e.g.
Rogstad et al. 1974 Nicholson, Bland-Hawthorn, Taylor 1992, see also
Fig.~16 in Schwarz 1985)
and follows the method described in Quillen et al.
(1992). The inclination and precession of the rings (representing gas orbits)
is given by continuous curves $\omega(r)$ and $\alpha (r)$, respectively (see
Fig. \ref{abbb3}).
A torque acting on an orbit with a circular velocity
$v_c(r)$ introduces a precession rate
$d \alpha/dt \sim \xi  v_c/r$. After a time $\Delta$$t$ one obtains
$\alpha(r) = \xi \Omega \Delta t  + \alpha_0$. Here
$\xi$ is given by the acting torque (see Appendix C)and $\Omega$=$v_c$/$r$.
We considered for our analysis models with constant traveling time
$\xi \Delta t$ and assume the molecular gas to be uniformly distributed. 
\\
{\bf Verification of the bar approach:}
To test how far our simplified bar models that follow the original approach 
of Telesco \& Decher (1988) are justified, we compare them to theoretical
calculations.
In our model continuous, smoothly varying curves of the ellipticities and 
position angles of the 3DRings ellipses were chosen under the
boundary condition that they do not cross each other. The resulting model
resembles the intensity map and velocity field of theoretical models
(e.g. {\em Model 001} of Athanassoula 1992b). 
The velocity field is shown in the rest-frame of
the rotating bar by subtracting a circular disk
model with constant intensity and a linearly increasing rotation curve
which results in the same angular velocity as the bar model at the co-rotation
radius of the bar model (Fig. \ref{abbb5}).
We also compare the density and velocity profile across
the dust lane. Due to shocks jumps in the density and
velocity profile occur at the position of the dust lanes (Fig. \ref{abbb6}). 
The density
profile shows a maximum at that position whereas the velocity has its
maximum upstream and adopts much smaller values downstream with respect
to the shock in the dust lane. As shown in Fig. \ref{abbb5} and \ref{abbb6}
the results of
3DRings are also qualitatively comparable to the ones of Athanassoula (1992b).
Therefore we are confident that our chosen bar approach is in
sufficient agreement with results of theoretical calculations and N-body 
simulations and very {\it well suited to search for bar signatures in the
observed $xyv$ cubes without having to calculate orbits in assumed or
inferred gravitational potentials}.

\newpage
\section{CAUSES FOR WARPS IN CIRCUM-NUCLEAR GAS DISKS}

Since a warped molecular gas disk in the circum-nuclear regions of
galaxies is a relatively new concept we give a more detailed description of
possible causes for warps in the area between a few to several 100 pc.
To generate a warp in a thin gas disk it has to be moved out of the plane
of the galaxy by an acting torque. Several mechanisms can account
for a torque in the central few hundred parsecs.
The most important ones are the
gas pressure of the ionization cone, radiation pressure of
a radio jet or a nuclear source, the gravitational 
forces of individual molecular cloud
complexes and an axisymmetric, non-spherical galactic potential 
(e.g. representing the stellar bulge of a galaxy).
\\
In general the torque $M$ can be expressed as

\begin{equation}
\label{eec1}
|\vec{M}| = |\vec{F} \times \vec{l}| = |\vec{F}| \cdot l \cdot
sin(\psi) \approx
p \cdot A \cdot l
\end{equation}

\noindent
For this  approximation we assume that the force $\vec{F}$ is
perpendicular to the lever arm $\vec{l}$ (i.e. $\psi$ = 90$^o$). 
In addition the force can be
expressed by a pressure $p$ onto the local disk area $A$ relating the
torque to a pressure gradient between the top and bottom side of
the disk. 
For the different mechanisms, we describe in the following approximations 
of the torques as well as the related observational quantities that are
required for this.
The results of the calculations are summarized in Tab. \ref{vv10}
and Tab. \ref{vv11}. Only the
torque by the gas pressure can sustain a warping of the disk
over several dynamical time scales. 
As a transient phenomenon a GMC above or below the disk
may induce a torque as well.

\subsection{Warp caused by an axisymmetric galactic potential}

The observed HI-Warps are explained by this mechanism (see review by
Binney 1992). The axisymmetric potential (Fig. \ref{abbc1} a)
of the halo applies a torque
to the HI gas disk evoking a warping of the HI disk. The derivation
of the torque is given in Goldstein (1980) and in Arnaboldi \& Sparke
(1994). The torque $M$ induced by a mass $m$ at a point ($r,\theta$)
with the inclination $\theta$ of the rotation axis of the orbit of the
mass $m$ relative to the rotation axis of the main system is here
defined as
$M \sim m \frac{\delta V}{\delta \theta}$
 (Stacey 1969).
Here $V$ is the axisymmetric potential. Arnaboldi \& Sparke (1994)
deduced in their analysis of orbits in polar rings in an axisymmetric
potential the torque far outside of the core of a slightly oblate
halo. For a potential with oblate symmetry (axial ratio $a=b\neq c$)
the torque is described by

\begin{equation}
\label{eec3}
M \sim m \pi G \rho_o \frac{a^2 (a^2 - c^2)}{3c^2} sin(2 \theta).
\end{equation}

\noindent
This relation is adequate to estimate the strength of an axisymmetric
potential outside the core radius. Inside the core radius of the
potential (halo) the torque varies with $\sim r^2$ (Sparke 1996).
The author also gives a relation between the precession rate and the torque:

\begin{equation}
\label{eec4}
M \sim m \dot{\Phi} sin(\theta) (r^2 \Omega(r)).
\end{equation}

\noindent
Here $\dot{\Phi}$ is the precession rate (in units of [rad/s]) 
and $\Omega(r)= \frac{v(r)}{r}$ is the angular velocity. 
Since these quantities are
known or used as input parameters for 3DRings (see Appendix~B),
the above relation can
be used to estimate the torque of the observed or 
modeled (with 3DRings) warp. 
We neglected the effects of dissipation, radial mass
transport (only circular orbits are used) and self-gravity within the
rings.
Both relations can only give rough estimates as the core size of the
axisymmetric potential is unknown or the precession rate $\dot{\Phi}$
(some fraction of $\frac{2\pi}{T}$; $T$ is the time of circulation
of a circular orbit) and the inclination angle $\theta$ have to be
determined from the model 3DRings. 
However, for a first approximation these estimates can be used as
reference values in order to determine whether a given mechanism 
is sufficient to account for the required torques.

\subsection{Torque imposed by a molecular cloud}

As a transient phenomenon a GMC above or below the disk
may induce a torque as well (Fig. \ref{abbc1} b).
The decomposition of the motions in the molecular gas in NGC~3227
(see section \ref{ww24} and Appendix~A) has shown that  in the
inner 300~pc molecular cloud complexes exist 
that do not participate in the circular motion of the underlying
molecular gas disk.
Therefore these complexes are likely not to be part of the gas disk 
but they are probably located above or below the disk.
In this case they interact gravitationally with the underlying gas
disk and can result in a warp. 
Under the assumptions that the mass $m_{GMC}$ of the GMC is
similar to the mass $m$ of the underlying gas disk segment and that
the GMC has a height above the disk of the order of its radius the
acting force between the GMC and the disk section
can simply be estimated via

\begin{equation}
\label{eec7}
F = G \frac{m_{GMC} \cdot m}{r^2}~~.
\end{equation}

\noindent
This then allows us to estimate the torque following equation \ref{eec1}. 
The mass of the
GMC can be estimated from the \CO line flux (see section \ref{ww28}). 

\subsection{Warp caused by gas pressure}

In active galaxies ionization cones ranging from the nucleus till a
distance of a few hundred parsecs are observed. These cones can be
regarded as parts of the NLR (narrow line region) for which the number
density $n$ and excitation temperature $T$ can be deduced from the
observed forbidden lines. The cones have partly large opening angles
and show in general a large inclination relative to the normal axis
on the disk. In many cases the cones can touch or even intersect the disk. 
Then there will be a pressure gradient
between the top and bottom part of the disk (Fig. \ref{abbc1} c).
Even if the molecular gas is clumpy the response of clumps to 
the pressure gradient imposed by the gas in the ionization cone
can be substantial (e.g. Eckart, Ageorges, Wild 1999) and due to
self-gravitative interaction (e.g. Lovelace 1998) the entire 
molecular disk will response.
In this case the gas pressure applies a force at a  distance $r$ from
the center onto the disk. 
Via a torque $\vec{M}$ this causes a change of the
angular momentum $\vec{L}$. This process can be stable over time-scales
which are long compared to the dynamical time-scales at the
corresponding radii. Therefore this scenario is very well suited to
cause and maintain warps in circum-nuclear molecular gas disks. 
Such a scenario has also been proposed by Quillen \& Bower (1999) for M~84.
To first order the gas pressure can be estimated from the 
equation of state for an ideal gas 

\begin{equation}
\label{eec8}
p = \frac{N}{V} \cdot k \cdot T = n \cdot k \cdot T
\end{equation}

\noindent
The number density $n$ can be derived from the particle number $N$ per 
volume $V$ and $k$ is the Boltzmann constant. 
The torque can then again be calculated following equation \ref{eec1}.
 The area $A$ is the interaction area
of the ionization cone with the galaxy disk and $l$, the lever arm,
is the distance to the center of this area.
For a purely geometrical dilution of the number density
it can be shown that the torque is independent of the
radial distance from the center.
The ionization cone and its influence on the
disk can therefore extend over a few 100~pc and thus provide a
continuous warping of the gas disk.

\subsection{Warp caused by radiation pressure}
\begin{figure}
\begin{center}
\psfig{file=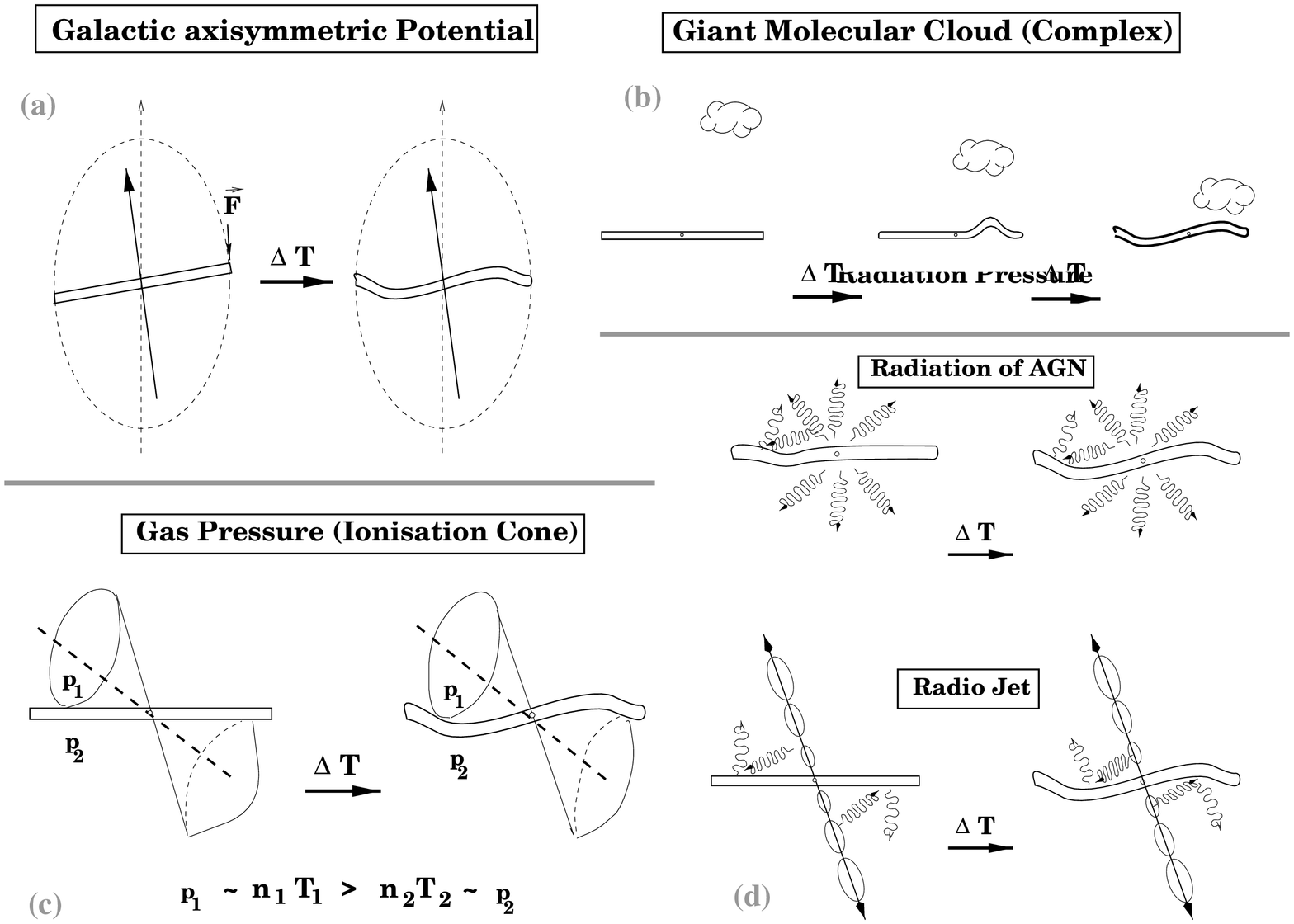,height=10.0cm,width=14.1cm,angle=-90.0}
\end{center}
\figcaption[]{
\label{abbc1}
(a) Influence of an axisymmetric potential onto a rotating gas disk. If the
rotating axis of the gas disk is misaligned to the figure axis of the
potential, this leads to a warping of the gas disk
after a time $\triangle T$.
\newline
(b) Influence of a molecular cloud onto a rotating gas disk. A molecular
cloud which is located above the gas disk applies a force onto the disk.
Within a dynamical time-scale a warp in the disk develops. 
Here we assume that the height of the cloud is of the order of its radial
distance to the nucleus. This time-scale over which the warp
develops is about equivalent to the orbital time scale of the 
gas in the disk at the
given radius as well as the time-scale in which the cloud is settling
into the gas disk. Dissipative interaction of the cloud with the disk
lead to the fact that this is only a transient phenomenon.
\newline
(c) Influence of an ionization cone onto a rotating gas disk. In this
scenario the ionization cone is in touch with the gas disk. Due to the
higher product of particle densities $n_1$ and temperatures $T_1$ 
in the cone a pressure gradient exists between the top and bottom 
part of the disk.
Because of the pressure difference $(p_1 - p_2)$ acting on the
interaction area a force and therefore a torque develops leading 
after a time $\triangle T$ to a warping of the gas disk.
\newline
(d) Influence of the radiation pressure onto a rotating gas disk.
\newline
{\it Top:} In the case of a circum-nuclear accretion disk a warping of
the disk can be caused by non-uniform illumination of the gas disk by the AGN
radiation field (UV, X-ray; Pringle 1997). 
\newline
{\it Bottom:} At larger distances the radio jet which lies out of the
plane of the gas disk can be regarded as a possible radiation source.
In this case non-uniform illumination of the gas disk is given such
that a warping mechanism - similar to the one described above - acts.}
\end{figure}

In the case of a 1~pc diameter circum-nuclear accretion disk a warping of
the disk can be caused by non-uniform illumination of the gas disk by the AGN
radiation field (UV, X-ray; Pringle 1996, 1997; Fig.
\ref{abbc1} d). This is due to the fact that independent of
the angle of incident photons are irradiated
perpendicular to the disk (Pringle 1996). The non-uniform illumination can be
evoked via a non-isotropic radiation field of the AGN or via a small 
instability in the disk that moves gas out of the plane (as shown here).
Due to extinction in the disk
this process can not act at large distances from the nucleus.
The radio jet, however, may represent a possibly strong radiation source.
If the jet is inclined with respect to the gas disk the jet components
can be quite close to the disk even at large radii from the nucleus
(Fig. \ref{abbc1} d).
Therefore we estimate the jet's possible contribution to the warping 
of the gas disk.
To calculate the radiation pressure of the jet its radiation power has
to be estimated first. We assume that the radiation flux density $S$ is
given by a power law $S = b \cdot \nu^{\alpha}$.
We assume synchrotron emission and $\alpha$$\sim$-1.
In order to obtain an upper limit to the torque induced by a jet
we also assume that the upper cut-off frequency is in the NIR 
to optical range as it is found for powerful jets (e.g. M87;
Meisenheimer et al. 1996). In the radio to
sub-mm range the disk can be regarded as being transparent. 
To estimate the total luminosity we used integration limits of 10
$\mu$m - 0.5 $\mu$m.
For isotropic radiation the jet luminosity is given via
$L = 4 \pi D^2 \int_{\nu_1}^{\nu_2} S d\nu
  = 4 \pi D^2 \int_{\nu_1}^{\nu_2} b \cdot \nu^{\alpha} d\nu$
Here $D$ is the distance to the source. 
The radiation
density $I$ onto a given area $A$ (seen under a given solid angle
from the jet) at distance $l$ can now be
calculated. The radiation pressure is then estimated via:

\begin{equation}
\label{eec14}
p = \gamma \frac{I}{c}.
\end{equation}

\noindent
Here $\gamma$ equals $1$ in the cause of a black body, $2$ for an ideal
reflecting body and $c$ is the speed of light. Again equation \ref{eec1} can
be used to obtain an upper limit of the torque due to radiation pressure
from the jet. 

\newpage

\notetoeditor{The following captions and figures belong to the appendices.}


\clearpage

\clearpage

\clearpage

\clearpage
\notetoeditor{The following two tables should be one (Tab.8);
         table caption is incomplete in the processed file (ps-file)}

\begin{table}[htb]
\caption{
\label{vv10} 
Causes for the warp in NGC~3227}
\begin{center}
\begin{tabular}{lllr}\hline \hline
Cause & Property & & Value \\ \hline
Estimated via & & &\\
{\bf 3DRings}$^a$   & {\bf torque}$^1$         & $M$ & {\bf 6.4$\times$10$^{46}$ Nm  }  \\
                & molecular gas mass           & $m$ & 2.0$\times$10$^6$ M$_{\odot}$  \\ 
                & radius                       & $r$ & 7.7$\times$10$^{17}$ m   \\ 
                & velocity                     & $v(r)$ & 60 km s$^{-1}$            \\
                & orbital time scale           & $\dot\Phi^{-1}$ & 1.8$\times$10$^{13}$ s      \\ \hline

{\bf Potential}$^b$ & {\bf torque}$^2$         & $M$ &{\bf 3.0$\times$10$^{10}$ Nm }   \\
                & molecular gas mass           & $m$ & 2.0$\times$10$^6$ M$_{\odot}$  \\ 
                & volume                       & $V_o$ & 2.9$\times$10$^{36}$ m$^3$    \\ 
                & total mass                   & $m_o=\rho_o V_o$ & 3.4$\times$10$^7$ M$_{\odot}$  \\ \hline
{\bf GMC}$^c$       & {\bf torque }$^3$         & $M$ &{\bf 1.4$\times$10$^{46}$ Nm  }  \\
                & cloud mass                    & $m_{GMC}$ & 3.7$\times$10$^6$ M$_{\odot}$  \\ 
                & force                         & $F$ & 5.4$\times$10$^{26}$ N    \\
                & lever arm                     & $l=r$ & 2.6$\times$10$^{18}$ m    \\ \hline
\hline 
\end{tabular}
\end{center}
See caption in Tab. \ref{vv11}
\end{table}

\begin{table}[htb]
\caption{
\label{vv11}
Causes for the warp in NGC~3227}
\begin{center}
\begin{tabular}{lllr}\hline \hline
{\bf Gas pressure}$^d$  & {\bf torque }$^4$     & $M$ & {\bf 1.2$\times$10$^{46}$ Nm }   \\
                & particle density              & $\frac{N}{V}$ & 1000 cm$^{-3}$  \\ 
                & temperature                   & $T$ & 1.4$\times$10$^5$ K       \\ 
                & gas pressure                  & $p$ & 1.9$\times$10$^{-9}$ Nm$^{-2}$  \\
                & area                          & $A$ & 4.7$\times$10$^{36}$ m$^2$  \\ 
                & lever arm                     & $l$ & 1.3$\times$10$^{18}$ m    \\ \hline

{\bf Radiation pressure}$^e$ & {\bf torque}$^5$ & $M$ & {\bf 1.2$\times$10$^{39}$ Nm  } \\
                & spectral index                & $\alpha$ & $\sim$ -0.9         \\
                & constant                      & $b$ & 8.4$\times$10$^{-19}$ Wm$^{-2}$  \\
                & luminosity                    & $L$ & 8.4$\times$10$^{30}$ W   \\ 
                & intensity                     & $I$ & 4.0$\times$10$^{-7}$ Wm$^{-2}$ \\ 
                & radiation pressure            & $p$ & 1.3$\times$10$^{-15}$ Nm$^{-2}$  \\
                & area                          & $A$ & 6.7$\times$10$^{35}$ m$^2$ \\ 
                & lever arm                     & $l$ & 1.3$\times$10$^{18}$ m   \\ 
\hline \hline
\end{tabular}
\end{center}
$^a$ eq. \ref{eec4};
~~~$^b$ eq. \ref{eec3};~~~$^c$ eq. \ref{eec1} and \ref{eec7};
$^d$ eq. \ref{eec1} and \ref{eec8};~~~$^e$ eq. \ref{eec1} and \ref{eec14}\\
$^1$ from the parameters of 3DRings for a nuclear disk and a radius of
     25~pc.\\
$^2$ from the nuclear molecular gas mass and the dynamical mass in a
     radius of 25~pc. (See section \ref{ww28}.) We used $M \sim m \pi G \rho_o$
     to get the order of magnitude.\\
$^3$ from the molecular gas mass of the redshifted knot and its	
     distance (84~pc) from the nuclear  gas disk (see
     section \ref{ww28}).\\
$^4$ particle density and temperature of the ionization cone from 
     Gonzal\'{e}s-Delgado \& P\'{e}rez (1997),  area ($\sim$(0.7''
     $\times$ 1.0'')) and distance ($\sim$ 0.5'') from HST [O~III] map 
     (Schmitt \& Kinney 1996).\\
$^5$ spectral index (for calculation we assumed $\alpha$ = -1.0) and
     constant $b$ as well as area (about 0.5'' $\times$ 0.2'') 
     to match the radio jet (Mundell et al. 1992b) at a distance of 42~pc.
\end{table}

\end{document}